\documentclass[manuscript=article,longbibliography]{achemso}
\usepackage{caption}
\usepackage{subcaption}
\usepackage{graphicx}
\usepackage{bm}
\usepackage{amsmath}
\usepackage{color}
\usepackage[linktocpage,colorlinks=true,linkcolor=blue,citecolor=blue,breaklinks=true,urlcolor=blue]{hyperref}
\usepackage{lineno}
\usepackage[normalem]{ulem}
\linenumbers

\DeclareUnicodeCharacter{2212}{-}

\title{Active dislocations and topological traps govern dynamics of spiraling filamentous cyanobacteria} 

\author{Xingting Gong}
\affiliation{Department of Applied Physics, Stanford University}
\author{Manu Prakash}
\email{manup@stanford.edu}
\affiliation{Department of Bioengineering, Stanford University}
\alsoaffiliation{Department of Biology (courtesy), Stanford University}
\alsoaffiliation{Department of Oceans (courtesy), Stanford University}
\alsoaffiliation{Woods Institute for the Environment, Stanford University}
\alsoaffiliation{Chan Zuckerberg Biohub, San Francisco}
\begin{document}
\nolinenumbers
\newpage

\begin{abstract}
Activity can organize matter in unique configurations inaccessible to equilibrium systems, including a sundry of spiraling shapes seen in nature that range from galaxies to living tissues to fossilized stromatolites. How these dynamic yet stable patterns form in motile active systems that span a range of length and time scales remains an open question. Here we study the collective gliding dynamics of ultra-long filamentous cyanobacteria confined in two dimensions and present the discovery of an emergent pattern we call ``active spirals". Individual filaments in the spiral bulk remain confluent due to adhesion forces and exhibit reversible gliding motility. Thus individual filaments undergo bidirectional movement and the spiral object as a whole has no fixed vorticity. Using single filament tracking, we discover that spirals permit the radial flux of material as filaments shear past one another. We demonstrate that these rearrangements can be entirely described by topological rules of interaction between filaments tips. We thus reduce the dynamics of a spiral to a set of active dislocations (corresponding to the filament tips) on a polar coordinate lattice and show that we can reproduce and predict the material flux in the system. Finally, we present a discovery of a novel topological trap present in these spirals, and is induced purely by the geometric chirality of long winding filaments with winding number greater than zero. A topological trap creates boundaries in the spiral across which material cannot flow, leading to persistent structures that are topologically locked for the lifetime of the system. The emergent mechanics of active spirals presented here sheds light on the critical role of adhesion forces, activity and geometry in the formation of long-term, stable, yet dynamic active patterns.
\end{abstract}

\maketitle

\newpage
\section*{Introduction}

Active matter is a rich field with the ability to inspire new paradigms in design and engineering due to its ability 
to collectively organize from the ``bottom-up''. In active materials, complex and dynamic behaviors emerge naturally from local rules of interaction between energy-consuming particles, whereas systems in equilibrium evolve toward a limited set of states determined by externally applied boundary conditions. Despite the rich phenomena accessible to these systems, it is difficult to disentangle the key rules of interaction that govern the emergent properties of interest \cite{Falk2021control}, and as such, engineering applications inspired by active materials are still in their infancy \cite{ross2019controlling}. 

One active structure of intrigue is an ``active spiral''. At the smallest length scales, these spiraling architectures can be seen in DNA torroids \cite{hud2001cryoelectron} for efficient genome packing, spiraling springs such as ESCRT filaments \cite{chiaruttini2015relaxation} for membrane cessation, collective motility of malaria parasites \cite{Patra2022}, or spiraling tissue structures in corneal epithelium \cite{lobo2016self}. In the non-living world, the most classical example consists of spiraling galaxies that form over billions of years at astronomical length scales\cite{sellwood2022spirals}. Each above is an example of a spiral structure that self-assembles from simple rules arising from the interplay between activity and geometry \cite{kruse2005generic}. Furthermore, many of these dynamic structures are incredibly stable over long time scales -- some even immortalized as micro-fossils of filamentous bacteria from the Precambrian era \cite{schopf2012fossil} -- while the individual components comprising the structure might be in flux, akin to the assembly of a spindle \cite{dumont2014emergent}. What provides this degree of stability and robustness to active patterns in a highly dynamic environment remains an open question. 

A challenge in studying active systems at small length scales is the resolution at which micro-states can be imaged. Many systems such as reconstituted protein mixtures or in-vivo motor-filament suspensions \cite{ross2019controlling} cannot be resolved at the resolution of their constituent particles (single filaments and motor molecules). Usually, coarse grained parameters such as local nematic order is measured, which creates an incomplete picture of particle-particle interactions. \cite{keber2014topology,sanchez2012spontaneous,huber2018emergence,schaller2010polar,zhang2021spatiotemporal,ross2019controlling,balchand2016using}. Machine learning is being used to tackle this challenge, but it again suffers from interpretability of the local interaction rules that critically define the properties of these active materials \cite{schmitt2023zyxin}. 

Here we present the discovery of a unique active spiral formed by a species of gliding filamentous cyanobacteria, \textit{Oscillatoria sp.}, where every individual filament is able to be resolved and tracked. When a dense clump of filaments is deposited on an agarose gel and compressed with a cover-slip, we observe the formation of beautiful two-dimensional spirals consisting of long, interwoven filaments (Fig. \ref{Fig1}, Movie S1). Once formed, these spirals persist stably for many hours until cell death, and consist of anywhere from a single to hundreds of filaments. The aforementioned motility of the filaments, refereed to as gliding, allows these filaments to self-propel across a surface without the aid of flagella or cilia, with motion directed along the long-axis of the organism \cite{halfen1971gliding}. As filaments glide, collisions with other filaments cause them to locally align with their neighbors, and as such the system can be considered a dry active nematic \cite{doostmohammadi2018active}. 



In contrast to other active nematics that favor half-integer defects \cite{dell2018growing,sanchez2012spontaneous,doostmohammadi2018active},  \textit{Oscillatoria sp.} consistently form vortex-like spirals due to their flexibility and long lengths \cite{isele2015self,duman2018collective, kurjahn2023quantifying, faluweki2023active}. It has been shown that self-propelled, flexible filaments can curl into spirals due to self-interaction \cite{isele2015self, duman2018collective}. For multiple filaments, this acts as a key nucleation event which then recruits other neighboring filaments to wrap around it, forming a large spiral structure (Fig. \ref{Fig1}). Although circulating vortex structures have  been previously seen in gliding filament assays using microtubules and actin, in both cases the resulting vortex diameters have been larger than their constituent filaments, such that individual filaments do not wind multiple times inside the structure \cite{sumino2012large, sciortino2021pattern}. In contrast, an \textit{Oscillatoria sp.} filament consists of a row of cells that remain adjoined at the cell walls, and can be as short as a few microns to as long as a few millimeters, corresponding to a few to several hundreds of cells in a single filament. The large aspect ratio of  \textit{Oscillatoria sp.} filaments allows for multiple windings and the formation of true spirals that, as we will show in this manuscript, have unique topological properties. Though filament elasticity has been shown to play a key role in the class of patterns \cite{joshi2019interplay} in active namatics, extremely long aspect ratios (of order 1:1000) such as the ones observed here have been less explored \cite{copenhagen2021topological}.

Apart from their large aspect ratios, filamentous cyanobacteria have other properties that make them a nice system in which to study collective motion. Filament growth is minimal under uniform low-light conditions (our chosen experimental condition), such that motility is isolated as the dominant driver of spiral dynamics (Fig. \ref{lengths_vs_time}). When gliding on an elastic substrate, filaments exhibit spontaneous directional reversals that are coordinated at high speeds across the entire length of the multicellular organisms (Fig. \ref{Fig2}D) \cite{bryant2006molecular}. This remarkable coordinated reversal adds an element of stochasticity to filament dynamics and the entangled spirals do not have a fixed vorticity. In addition, the filaments are enclosed in a sheath of mucus which is secreted as they glide\cite{bryant2006molecular}. These slime layers play an important role in the collective behavior of this organism, as it helps the filaments adhere to both the surfaces and to each other \cite{hoiczyk1998junctional, morales2019role, formosa2018role, kurjahn2023quantifying}. All these individual properties (substrate based gliding motility, long aspect ratio, directional reversal, particle-particle adhesion) give rise to a unique active system exhibiting a number of interesting and complex behaviors.
 
In the remainder of this paper, we first characterize the physical properties and dynamics of spiral structures through single filament tracking. Our tracking reveals that the material within a spiral is highly dynamic, with filaments able to interweave and swap position. Since the structure is strictly two-dimensional with no out-of-plane cross-overs allowed, we demonstrate that this material flux is governed only by rules of interactions between filament tips. We show that the entire system can be described by dislocation dynamics, where filament tips can be considered as active dislocations in a 2D polar lattice. Using this, we present a framework that reduces the collective motility of filaments to the dynamics at their tips. We show that it is possible to enumerate all the rules of tip-tip interactions on a polar coordinate lattice, which we then implement \textit{in silico} to compute how the degree of material flux within a spiral varies with reversal dynamics. Comparison of these results with experimental data strongly suggests the existence of biological feedback mechanisms that control the rate of filament reversal.

Finally, we conclude the manuscript with the discovery of a novel topological trap that prohibits the flux of material in a spiral. These traps arise naturally from the geometry of long filaments with winding number greater than zero, and  might provide an intriguing mechanism for the stability of active patterns over long time scales.  

\section*{Spiral dynamics}

We seed \textit{Oscillatoria sp.}  filaments on an agarose gel and confine the system to two-dimensions by compressing it with a coverslip. Over the next few hours, the system naturally goes from an entangled mass to a well ordered active nematic system purely in two dimensions (Fig. \ref{Fig1}, Movie S1). Of interest to us are the regions of spontaneous spiral formation, particularly spirals that exist in isolation (not touching neighboring streams of filaments). 

Spirals can vary in size from just a single filament that coils on itself, to several hundreds of filaments (see Fig. \ref{spiral_stats}, Movies S2-S7 for six different tracked spirals of varying size). Fig. \ref{Fig2}A depicts the largest tracked spiral with a total of 59 filaments, where each filament is given a unique coloring and tracked for a duration of 5 hours. During this time, the spiral has achieved a steady state size, where the mean size of its inner central area neither shrinks nor expands but oscillates about a steady state (Fig. \ref{Fig2}B). Filaments in this spiral span three orders of magnitude in length, ranging from the shortest being 7 $\mu$m to the longest being 2965 $\mu$m (Fig. \ref{Fig2}C). Growth rates for \textit{Oscillatoria} have previously been measured to be slow, with a doubling rate of over 3 days when incubated at 20$^\circ$C \cite{foy1983interaction}. Indeed, our data shows negligible changes in length of the tracked filaments over the course of our experiments which are imagined under low light conditions (see Fig. \ref{lengths_vs_time}). Hence, motility rather than growth accounts for the dynamics observed. 

Individual filaments are seen swirling in both clockwise and counterclockwise directions within a spiral, and are subject to spontaneous directional reversals. Fig. \ref{Fig2}D depicts two sample velocity traces, where the orange and blue curves in each panel correspond to the (linear) velocities of each of the filament tips, plotted separately. In these traces, a positive velocity denotes counter-clockwise (CCW) motion, whereas a negative value corresponds to clockwise (CW) motion. 
Reversal frequencies are measured to lie in the range $\lambda \in (0.2, 1.2) \text{min}^{-1}$, and appear to have a length dependence, with shorter filaments tending to reverse direction more frequently (up to 5 fold increase, Fig. \ref{Fig2}C). Cyanobacteria filaments exhibit phototaxic and photophobic behaviors \cite{wilde2017light,nultsch1983partial}, suggesting that an active mechanism exists to control the direction of filament movement. However, all of our timelapses are taken under uniform, low-light conditions, minimizing the effect of light-dependent directional biases in our experiments. With a few exceptions (Fig \ref{tips_moving_opp_dir} and Movie S8), directional reversals occur instantaneously across the filament length (or at least comparable to the frame rate of imaging). We also note that the instantaneous velocities of all filaments satisfy a symmetric bimodal distribution (Fig. \ref{Fig2}E), and on average half of the filaments move in one direction versus the other (Fig. \ref{Fig2}F). The spiral structure as a whole therefore exhibits no preference in direction. 

Our single filament tracking (see Section III of supplementary materials for details) reveals, somewhat surprisingly, that the flow of material is highly dynamic within a stable spiral. While the velocity of the filaments is primarily tangential, we find that filaments are able to interweave and traverse radially inward and outward. In the depicted kymograph of Fig. \ref{Fig2}G, we clearly see signatures of individual filament motion along the radial coordinate through a section at $\theta = 0$. Tracking the position of a single filament over time (black dashed line, Fig \ref{Fig2}G) allows us to visualize this material flow, where the pink filament traverses a radial distance of 52 $\mu$m over the course of five hours (Fig. \ref{Fig2}H, Movie S9). 

A final important feature of these spirals arises from the slime trails, which are secreted by \textit{Oscillatoria sp.} as they glide. This layer of slime helps filaments adhere to surfaces and to each other \cite{hoiczyk1998junctional, morales2019role, formosa2018role}. The presence of these adhesion forces is a central feature of active spirals presented here, as they enforce the bacterial layers to remain confluent and suppress any density fluctuations (regions of gaps with no filaments). By segmenting out the gaps within a spiral structure (Fig. \ref{energy_fig}D, \ref{gaps}), we make two observations: (1) gaps consistently account for less than 3\% of the spiral bulk, and (2) when gaps appear, they tend to localize around regions of unpaired tips, where the addition of a filament layer forces the neighboring filaments to bend around it. 

\section{Spiral shape fluctuations}

We now investigate further the details of elastic energy fluctuations of a spiral, to gain insight into a specific mechanism of spiral shape relaxation. A key property of active spirals is their stability -- once formed, they maintain a steady shape and size for long times, despite the activity of motile filaments in the bulk (Fig. \ref{Fig2}B, \ref{radial_slippage}). Fig. \ref{energy_fig}A depicts a spiral in which every segmented filament has been skeletonized, and the bending energy at each node is calculated using the formalism in Bergou et. al. \cite{bergou2008discrete}. 

As expected, the local curvature of the filaments increases around sites of unpaired tips (i.e. tips that are not ``neutralized'' by an adjacent tip in its lane, leading to a distortion of the filament layers surrounding them). In Fig. \ref{energy_fig}B, we plot the total bending energy (normalized by $\alpha$, the bending modulus) of 6 tracked spirals as a function of time. The first observation is that all spirals fluctuate stably about some average energy, indicating that they have achieved a steady state configuration and size. Secondly, the average values themselves show a reasonably large spread, suggesting that the filaments are flexible and that the energy required to bend them does not exceed the energy of adhesion required to hold the spiral structure in place. Thirdly, we observe two distinct shape fluctuation dynamics: spirals which fluctuate mildly about their average bending energy (5 of the 6 tracked spirals), and spirals that show significant oscillations (Fig. \ref{energy_fig}B, light green curve). 

The light green curve corresponds to the most tightly wound spiral, with a filament of length $\sim$3300 $\mu$m wrapped about a center of radius $\sim$20 $\mu$m (Fig. \ref{spiral_stats}E, Movie S6). The oscillation events in the spiral energy correspond to directional reversals (Fig. \ref{reversal_arrows}), where the innermost filament winds and unwinds, allowing the spiral center to shrink and expand. This behavior, however, is not observed for the other 5 spirals. Plotting reversal frequencies as a function of filament bending energy suggests that tightly wound filaments tend to reverse more frequently, but only above an energy threshold (Fig. \ref{reversal_vs_ebend}). Hence, filaments that are very tightly wound are capable of storing and releasing elastic energy by winding up to a point and then reversing direction. 

The remaining five spirals do not exhibit reversal-mediated shape oscillations, and instead fluctuate mildly about an average size. 
In fact, filaments at the very center of the spiral can continue winding ``inward'' (i.e. in toward the spiral center) for extended periods of time, not only without leading to a net decrease in the size of the spiral center, but even while experiencing increases in size (Fig. \ref{radial_slippage}, Movie S10). 

This paradoxical phenomena dictates the necessity of radial slippage, such that filament velocities are not strictly tangential. In the absence of reversal, there are two possible mechanisms for spiral shape relaxation. In the first mechanism, the spiral opens up like a torsional spring, experiencing radial slippage globally along its length as it continuously winds inward \cite{xie2014mechanics}. In the second mechanism, the spiral experiences local shape fluctuations due to the combined effects of tip dynamics and adhesion forces. Gaps (which localize around unpaired tips) appearing in the spiral bulk allow filaments in channels underneath it to expand outward, preventing the spiral center from shrinking even as an interior filament continues winding without reversal. To interrogate which of these two mechanisms explains our data, we correlate fluctuations of the boundary against locations of unpaired tips. 

Let $R_1(\theta)$ and $R_2(\theta)$ denote the inner and outer boundaries of the spiral, with both curves parametrized by the polar angle $\theta$ (Fig. \ref{energy_fig}E). Then $W(\theta) = R_2(\theta) - R_1(\theta)$ is the width of the spiral at a section $\theta$. Let $\Delta W(\theta)$ denote the change in width at $\theta$ between two consecutive frames of a spiral timelapse. For sections where  $|\Delta W(\theta)| > \Delta R$ or $|\Delta W(\theta)| < \Delta R$, we can ask whether an unpaired tip exists in some neighborhood $\theta \pm \frac{1}{2} \Delta \theta$. Performing this analysis for the tracked spirals, we see that regions of the boundary with displacements greater than $\Delta R = 2.2 \mu m$ (approximately $60\%$ of the filament width) are more consistently correlated with the presence of unpaired tips, and therefore gaps (Fig. \ref{energy_fig}F, red curves), than are regions of the boundary with displacements $|\Delta W(\theta)| < \Delta R$ (Fig. \ref{energy_fig}F, blue curves). Our data therefore strongly supports the second mechanism. If instead filaments behaved like torsional springs, we would expect displacements to occur globally and simultaneously along $\theta$, rather than preferentially in regions of unpaired tips. Spiral shape fluctuations are thus local rather than global phenomenon, and are governed by tip dynamics which generate gaps, and adhesion forces which close them. From the analysis of spiral shape fluctuations, tips begin to emerge as important areas of localized activity, an idea we explore further in proceeding sections by using an analogy between tips and edge dislocations in crystals. 

\subsection{Tips as active dislocations}

Unlike classical dislocation dynamics, out of equilibrium systems allows us to explore a rich set of phenomena due to the fact that individual tips are themselves ``active" \cite{marchetti2013hydrodynamics, doostmohammadi2018active, giomi2014defect, giomi2013defect}. In this section, we find a useful analogy between tips and edge dislocations in crystals (Fig. \ref{energy_fig}C, bottom panel). In a crystal lattice, an edge dislocation is a line defect where an additional plane of atoms exists within the crystal structure. Similarly, in a spiral, an unpaired tip marks the point where an additional layer of material is inserted between channels of other filaments. Taking a loop around an unpaired tip, we can compute the burgers vector which points radially (Fig. \ref{energy_fig}A inset). In a crystal, dislocations can move both perpendicular and parallel to the burgers vector, known as climb and glide dynamics, respectively. Due to the nature of a spiral where filament lengths remain contiguous, only climb dynamics are allowed (Movie S11). 

In a biophysical context, active climb dynamics have previously been explored in the context of bacterial cell wall growth \cite{Amir9833}. The work by Amir et. al. modeled the growth of bacterial cell walls using dislocations, where dislocations in the cylindrical peptidoglycan lattice represent potential sites of material insertion. By studying the dislocation dynamics on a cylindrical lattice, the authors were able to reproduce experimental phenomena such as the exponential growth of the bacterial cell wall, and the pinning of synthesis machinery at dislocation sites.

Unlike the bacterial cell wall where linked peptidoglycan layers exist at fixed intervals on a cylindrical lattice, filaments in a spiral do not have to break covalent bonds when inserting themselves in between their neighbors, needing only to overcome adhesion and bending forces as they push filaments in surrounding lanes out of the way. Thus, while elastic forces in Amir et. al. can cause pinning of motor activity at dislocation sites, we observe that bending energy in a spiral is not strong enough to stall the motion of filament tips. 

Recall that spirals exhibit material flux -- the ability for filaments to move radially and reorganize position. In two-dimensions, because filaments cannot glide over one another, the only way for two filaments to swap position is for their tips to interact: the leading tips of two filaments collide, allowing them to interweave and eventually change their radial ordering (Fig. \ref{energy_fig}C). Hence, material flux in a spiral is driven by active dislocations exhibiting climb dynamics (Movie S11). 
The reduced order description of an active material to its defects is not a new concept. Active nematics created by cytoskeletal suspensions are a well known example, in which $+\frac{1}{2}$ disclination defects are self-propelled and drive flows \cite{keber2014topology,doostmohammadi2018active,sanchez2012spontaneous,giomi2014defect,giomi2013defect,shendruk2017dancing}. 
Furthermore, $\pm \frac{1}{2}$ disclination pairs can spontaneously nucleate and annihilate, all the while keeping the total topological charge in the system constant. The analogy to nucleation and annihilation events in a spiral is the ability of tips to pair and un-pair in the same channel (Fig. \ref{energy_fig}C, Movie S11), effectively annihilating and creating dislocation pairs of opposite burgers vectors.  Hence, the total number of dislocations at any given time is not constant, but is bounded by the total number of filament tips. 

We have shown so far that spirals are composed of layers of interwoven filaments which remain confluent -- the spiral bulk remains uniformly dense, with no density fluctuations. Within the spiral bulk, dislocation dynamics -- i.e., activity at the tips -- mediate both spiral shape fluctuations and material flux, and are therefore a useful reduced order description. In the proceeding sections, we model a spiral on a polar-coordinate lattice and show that we can precisely enumerate all the possible rules of tip interactions. Endowed with these rules, we can then predict the flux of material in a spiral as a function of reversal frequency. We focus on spiral dynamics below the threshold at which energy-dependent reversals become important. In fact, we go one step further and work in a limit where reversal frequencies are not modulated by external stresses, and aim to understand how a simple stochastic reversal process affects and reproduces the material flux observed in experiment.

It has been shown that rod-shaped bacteria are able to control their collective formations by tuning their reversal frequencies \cite{thutupalli2015directional, liu2019self}. The collective behavior of gliding, elongated bacteria is dominated by nematic alignment through mechanical collision and by reversal dynamics, which dictates the run-lengths that organisms can make before switching direction. Similar concepts apply to the geometry of a spiral, where reversal frequency affects the movement of material in the system by either enhancing or decreasing the frequency of tip interactions. Hence, reversal frequency can be considered a parameter to optimize the search of space, which bacteria are known to regulate in response to environmental cues such as light and the availability of nutrients \cite{grossmann2016diffusion, wu2009periodic, thutupalli2015directional}. The question of how reversal dynamics affects the internal organization of material is therefore interesting from both an ecological and geometric perspective.

\section*{Dislocations on polar grids as a reduced order model for tip interactions}

Material flux in a spiral is governed by how tips interact with other tips within the densely packed spiral bulk. We show in this section that it is possible to enumerate all the tip-interaction rules to predict material flux in a spiral. Moving forward, we denote the leading tip of a filament with a ``+'' symbol and trailing tips as ``--''. As the filament moves, the leading tip ``adds'' material as it passes through a cross section at $\theta$, while the trailing tip ``subtracts'' material (Fig. \ref{sim_material_flux}A). 

In addition, because we are concerned only with the ordering of material in a spiral, without loss of generality we fix the inner boundary of the spiral and \textit{allow only the upper boundary to fluctuate}, with the radius of the inner boundary set to the average radius of the steady state spiral center. This simplifying assumption allows us to position the channels in which tips can move at discrete radii located at intervals set by the filament width (Fig. \ref{sim_material_flux}A). As tips move in circular channels, they can interact via collision events or adhesion forces that cause them to move into other lanes.

Because of the fixed lower boundary in our model, when two tips collide there is a 50\% chance for either of the plus tips to get bumped up a channel (that neither tip is allowed to move $\textit{down}$ a channel does not affect the topological ordering of filaments). To ensure that the lengths of each filament remain constant even as tips hop between lanes, we update the leading tips according to an average \textit{linear} velocity of $1 \ \mu$m/s, and update the trailing tips only when doing so does not cause the filament length to increase (see Supplementary Information for implementation details).

We now enumerate all the possible rules of tip interactions on a polar lattice. Let us first consider collision events between tips moving in opposite directions. Since every tip is endowed with type (plus or minus) and direction (CW or CCW), there are four possible flavors of tips: $\{(+, \text{CCW}), (+, \text{CW}), (-, \text{CCW}), (-, \text{CW})\}$. Given these flavors, there are four unique pairs of colliding tips moving in \textit{opposite} directions: $\{(+, \text{CCW}), (+, \text{CW})\}$, \\$\{(+, \text{CCW}), (-, \text{CW})\}$, $\{(+, \text{CW}), (-, \text{CCW})\}$, and $\{(-, \text{CCW}), (-, \text{CW})\}$. The first of these pairs corresponds to a collision between the plus tips of two oppositely moving filaments, as depicted in the first row of Fig. \ref{sim_material_flux}B,i. For such an event, it is equally likely for either the CW-moving or CCW-moving tip to get pushed to the outside of the other. 

Collisions between the following two pairs $\{(+, \text{CCW}), (-, \text{CW})\}$, $\{(+, \text{CW}), (-, \text{CCW})\}$ are mirror reflections of each other, and for simplicity we depict one of them in the second row of Fig. \ref{sim_material_flux}B,i. Note that the collision between oppositely moving tips of opposite type is constrained by geometry, and the result of the collision is therefore deterministic -- the minus tip must be pushed into the outer channel. 
Collision events between the final pair $\{(-, \text{CCW}), (-, \text{CW})\}$ is geometrically impossible to achieve in 2D, for it would require the two filament bodies to intersect. Thus, Fig. \ref{sim_material_flux}B,i depicts all collision events between pairs of oppositely moving tips. 

Now let us consider collisions between pairs of tips moving in the \textit{same} direction. There are six possibilities: $\{(+, \text{CW}), (+, \text{CW})\}$, $\{(+, \text{CW}), (-, \text{CW})\}$, $\{(-, \text{CW}), (-, \text{CW})\}$, and their \text{CCW} mirror reflections. Schematics of these collision events are shown in Fig. \ref{sim_material_flux}B,ii. In order for tips moving in the same direction to collide, they must be moving at different speeds (the double arrow in Fig. \ref{sim_material_flux}B,ii labels the faster moving tip). In both our simulation and in experiment, the instantaneous speeds of individual tips is not a constant, and hence it is possible to achieve collisions between tips moving in the same direction (see SI for details). As with Fig. \ref{sim_material_flux}B,i, the top row of Fig. \ref{sim_material_flux}B,ii is the only stochastic event, whereas the bottom row is deterministic due to geometric constraints.

Finally, we consider tip interaction rules due to adhesion forces. The presence of adhesion between neighboring filaments enforces the closure of gaps that appear within a spiral due to the motion of trailing tips that subtract material as they move (Fig. \ref{energy_fig}D). The left-hand side of Fig. \ref{sim_material_flux}B,iii shows what happens when a minus tip in a lower channel passes tips moving in upper channels. When faster moving tips move over (on top of) a minus tip, this corresponds to encountering the beginning of an edge dislocation, which causes the tips to move up a channel. Conversely, when slower moving tips slide behind a minus tip, this corresponds to a disappearing layer of material, and so tips merge down one lane. At no point are gaps created in the lattice during this process. A complementary set of rules for plus tips are listed on the right-hand side of Fig. \ref{sim_material_flux}B,iii. 

We have thus enumerated all possible rules of interaction between moving tips in a polar coordinate grid. In the next section, we implement these rules \textit{in silico} to understand how the internal filament organization within a spiral varies with reversal frequency.

 \section*{Material organization vs. reversal frequency}

To gain intuition for how reversal rates affect filament organization in a spiral, it is useful to think of the extremes of infinite and zero reversal frequency. The former case will lead to filaments oscillating in place, and it is obvious that material flux will be minimal. The case of zero reversal frequency is more interesting: at long times, the spiral architecture will evolve to a stable configuration that cannot support any new filament swaps even though filaments are actively moving. The schematic in Fig. \ref{sim_material_flux}C depicts what happens when two oppositely moving filaments begin to interweave due to a tip-collision event. Once the inner and outer filaments swap position, the two filaments cannot exchange again without reversal, for the filaments tips are now topologically prohibited from colliding and interweaving (Fig. \ref{sim_material_flux}C, green circle). In the absence of reversal, the set of allowable tip interactions has shrunk, such that only lane merges become possible and tip-collisions are excluded. We thus posit that an optimal reversal frequency might exist that maximizes material flux in a spiral.

Figs. \ref{sim_material_flux}D,E show the results of our simulations (modelled after the largest tracked spiral shown in Fig. \ref{Fig2}A, see also Movie S12) at four different reversal frequencies $\lambda$, all separated by an order of magnitude. We can measure the amount of material flux in the system by representing the spiral at each time point as a connectivity graph, where each node is a filament and an edge between nodes signify that the filaments are touching in adjacent channels (Fig \ref{sim_material_flux}F). Specifically (though not shown in the figure), whether a filament is an inside or outside neighbor is recorded separately, such that filament swaps are detected. The hamming distance is then computed between the adjacency matrix of a graph at time $t$ with the graph at time $0$. Figure \ref{sim_material_flux}G plots the hamming distances for experiment and simulation, which is repeated in triplicates for each value of $\lambda$. The average hamming distance reached is then plotted in Fig. \ref{sim_material_flux}H, which shows that the lowest reversal frequency ($\lambda = 0.0006 \ \text{min}^{-1}$) exhibits the least amount of material exchange, congruent with our intuition. At $\lambda = 0.6 \ \text{min}^{-1}$ we observe that the slope of the curve has changed sign. For $\lambda \to \infty$ we expect the amount of material flux to approach 0, and so the fact that the curve in Fig. \ref{sim_material_flux}H reaches a maximum and begins to decline is again congruent with our intuition.

By generalizing the schematic in Fig. \ref{sim_material_flux}C to an arbitrary number of filaments, one can prove that in the absence of reversal, a spiral will naturally phase separate into two bands: one that is clockwise-moving, and another that is counter-clockwise moving. If we were to compute the correlation in direction between filaments as a function of distance (excluding self-interactions between filaments with winding number $> 0$), we would expect in the perfectly phase-separated case a correlation that equals +1 for small inter-filament distances and -1 for large inter-filament distances. The exact location at which this function hits zero depends on the width of each band. When reversal frequency is nonzero, the spiral will no longer phase separate into two distinct bands, and we expect the direction correlation function to reflect this change. 

Fig. \ref{sim_dir_corr} depicts the direction correlation graphs for both experiment and simulation (as well as the corresponding kymographs, where each filament is now labelled by velocity). As expected, the simulation with the smallest reversal frequency has the clearest correlation trend that decreases from +1 to -1 (Fig. \ref{sim_dir_corr}E). Surprisingly, however, the simulation that most closely resembles the experimental result has a value of reversal frequency that is two orders of magnitude higher than what is measured experimentally (Fig. \ref{sim_dir_corr}D). The distribution of average reversal frequency (Fig. \ref{Fig2}C, blue) has a range of $(0.1 \text{min}^{-1}, 1.39 \text{min}^{-1})$ and mean 0.66 min$^{-1}$. Already at a rate of 0.06 min$^{-1}$, the signature of positive correlations at short lengths and negative at long lengths is wiped out (Fig. \ref{sim_dir_corr}C). 

We posited that perhaps a simple poisson process with constant reversal frequency for all filaments was too simple to reproduce the observed correlations. Fig. \ref{Fig2}C,blue shows that there is a length dependence for average reversal rates. In addition, plotting the distribution of dwell-times (times in between reversal) for all filaments reveals a heavy-tail, suggesting that filaments have an enhanced probability of making extremely long-runs before reversal (Fig. \ref{heavytail_dwelltime}). This is further reflected in the histograms of instantaneous velocities broken down by individual filaments, some of which exhibit significant skew in one direction versus another (Fig. \ref{063030_spd_histograms}). We thus ran two more simulations where we implemented length-dependent reversal rates in one, and sampled directly from the dwell-time distribution in the other. Neither simulations were able to reproduce the directional correlations, providing strong evidence that self-organization based on kinematics alone does not appropriately account for all interaction rules (Fig. \ref{LDR_dwell_dir_corr}). Biological feedback mechanisms which respond to bending or shear forces from neighboring filaments, for example, could be an important regulatory mechanism by which filaments modify their direction of motion \cite{dinet2021linking}. Our simple reduced-order model suggests that these second-order effects are important, and coupling elasticity and mechanical stress to reversal dynamics is an interesting avenue of future work.

\section*{Geometric chirality and topological traps}

We conclude this manuscript with an interesting discovery that arises from the geometry of long, wound filaments. 
In the previous sections, we explored how reversal dynamics affects the degree of material reorganization in a spiral. In this section, we discuss an important property which can place limits on material movement, a property we call geometric chirality. 

Fig. \ref{top_trap}B and \ref{top_trap}E show the kymographs of filament position for the two tracked spirals in Figs. \ref{top_trap}A,D. Most notably, tracking the motion of the dark red spiral in Fig. \ref{top_trap}B over time shows that it nearly weaves all the way from the inside to the outside. On the other hand, the kymograph in Fig. \ref{top_trap}E shows distinct boundaries across which material does not appear to cross (black dashed lines). Compared to the second spiral, material in the first is more fluid and able to traverse the entire bulk. To understand if this occurred by chance or due to an underlying mechanism, we came across an interesting observation. Tracing the length of the bright red filament in Fig. \ref{top_trap}D (which forms the outermost barrier at $R_3$) from its inside tip to its outside tip shows that it winds geometrically in the counter-clockwise direction (Fig. \ref{07022020-fil9_fil11}). Here we define this filament to have CCW \textit{geometric chirality}. Notably, tracing the length of any multiply-wound filament at $R > R_3$ shows that they all have clockwise geometric chiralities.

What happens when two filaments of opposite geometric chirality are wrapped around each other? In this scenario, the two filaments are topologically \textit{locked} from inter-digitating. Fig. \ref{top_trap}H shows a cartoon schematic of this situation, where the two colliding tips (the outer tip of the inner filament, and the inner tip of the outer filament) are labelled. When a collision event happens, the purple tip remains on the outside of the green tip, preventing the filaments from interweaving. This is true no matter what direction the two filaments are gliding. Two adjacent filaments of opposite geometric chirality therefore form a topological trap, preventing the flux of material through them.

In Figs. \ref{top_trap}C,F we now label the filaments of each spiral by geometric chirality (the unlabeled filaments have undefined geometric chiralities due to the fact that the winding number is less than one). As expected, the spiral in the upper panel has uniform CCW-wound filaments, and material is free to move throughout the bulk of the entire spiral. The second spiral, on the other hand, exhibits alternating regions of CW and CCW wound filaments. The boundaries of these regions correspond exactly to the radial distances across which material cannot flow (see Movie S13). Geometric chirality can thus limit material flux altogether -- two adjacent filaments of opposite geometric chirality give rise to a barrier across which material cannot enter or escape, and this so-called topological trap arises naturally from the geometry of filaments with winding number $> 1$.  
We note for the sake of completeness that the spiral in Fig. \ref{Fig2}, after which our simulations were modelled, do not have topological traps within the spiral bulk, and as such do not impact the results in Figs. \ref{sim_material_flux} and \ref{sim_dir_corr}. We view these traps as not only an interesting topological property that arises naturally from the geometry of long and wound filaments, but also a mechanism by which to program the internal organization of active spirals.

\section*{Discussion}

Our current work focuses on two-dimensional entangled spirals that form naturally in soft confinement. We explicitly choose to study structures where no material is entering or exiting the spiral in order to understand both the dynamics and stability of these patterns.
By utilizing dim and uniform lighting conditions, we avoid any significant growth in these filaments during observations as well as any phototaxic behaviors that would bias motility patterns. 

Our single filament tracking reveals the rich dynamics of individual filaments that undergo radial flux and directional reversals, all the while the overall spiral structure remains stable.
In addition, the presence of adhesion forces ensure that the spiral bulk remains as a single confluent layer, allowing us to model a spiral as a system of active dislocations moving in a polar coordinate lattice. Our simple model of stochastic reversals suggests the presence of force feedback mechanisms at play which may regulate bacterial reversal dynamics. Cyanobacteria filaments can possibly sense mechanical parameters such as forces, confinement pressure or bending beyond a certain radius of curvature. Coupling energetics to reversal events to build a model that captures the mechanobiological link between induced stress in filaments and their dynamics is an important avenue of further research.

Other avenues of future work include exploring how this entangled structure can respond to external cues. Such experiments include using light as a parameter to perturb the formation of the spiral structures by exploiting phototaxis. In addition, sonication techniques can be used to tune average filament lengths (filaments that mechanically fracture continue to live as two separate filaments), potentially giving rise to different patterns. 

Finally, we conclude our observations with the discovery of a topological trap -- a characteristic unique to long filaments capable of winding more than once. No matter their direction of motion, two filaments of opposite geometric chirality will never be able to interweave, forming a barrier to material flux. Reversal frequency and geometric chirality are therefore two parameters by which one can control the organization of material within a spiral. We hope that the concept of a topological trap stabilizing a biological pattern will find broad utility in active matter dynamics. Taken together, our results not only shed light on an interesting geometric and biological system, but moves the needle ever so slightly towards the ultimate goal of engineering and controlling active materials. 



\section*{Acknowledgments}

We thank all members of PrakashLab for helpful discussions. We particularly thanks Haripriya Mukundarajan, for engagement in early phases of this work. X.G and M.P conceptualized this work, designed the experiments and analysis techniques used. X.G conducted all the experiment after initial support from M.P. X.G developed all the code for data analysis and simulations. All authors contributed to interpreting data, and writing the manuscript.  The work was supported by  a HHMI Faculty Fellowship (M.P.), Bio-Hub Investigator Fellowship (M.P.), Schmidt Innovation Fellowship (M.P.), Moore 
Foundation Research Grant (M.P.) and NSF CCC (DBI1548297 (M.P.)). 

\newpage

\begin{figure*}
\includegraphics[width=\textwidth]{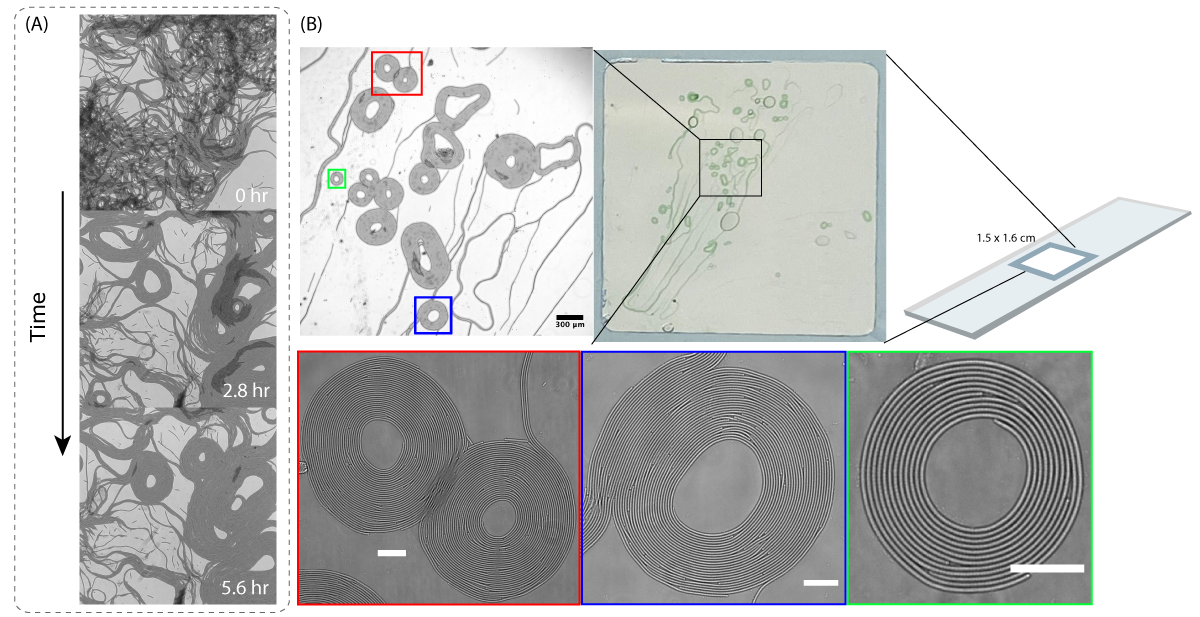}
\caption{\label{Fig1} 
Active spirals formed from filamentous \textit{Oscillatoria sp}. All unlabelled scale bars show 50$\mu$m. \textbf{(a)} Evolution of an intial clump of filaments placed on top of an agarose gel (total surface area 1.5 $\times$ 1.6 cm and volume 65$\mu$L) and compressed with a coverslip. Over time, filaments spread out into locally aligned 2D mats. Note the emergence of spiral structures composed of wound filaments. \textbf{(b)} The upper row depicts progressively zoomed in regions of the agarose gel. The second row shows close-ups of various spirals. All images are taken over 48 hours after initial plating. 
}
\end{figure*}

\begin{figure*}
\includegraphics[width=\textwidth]{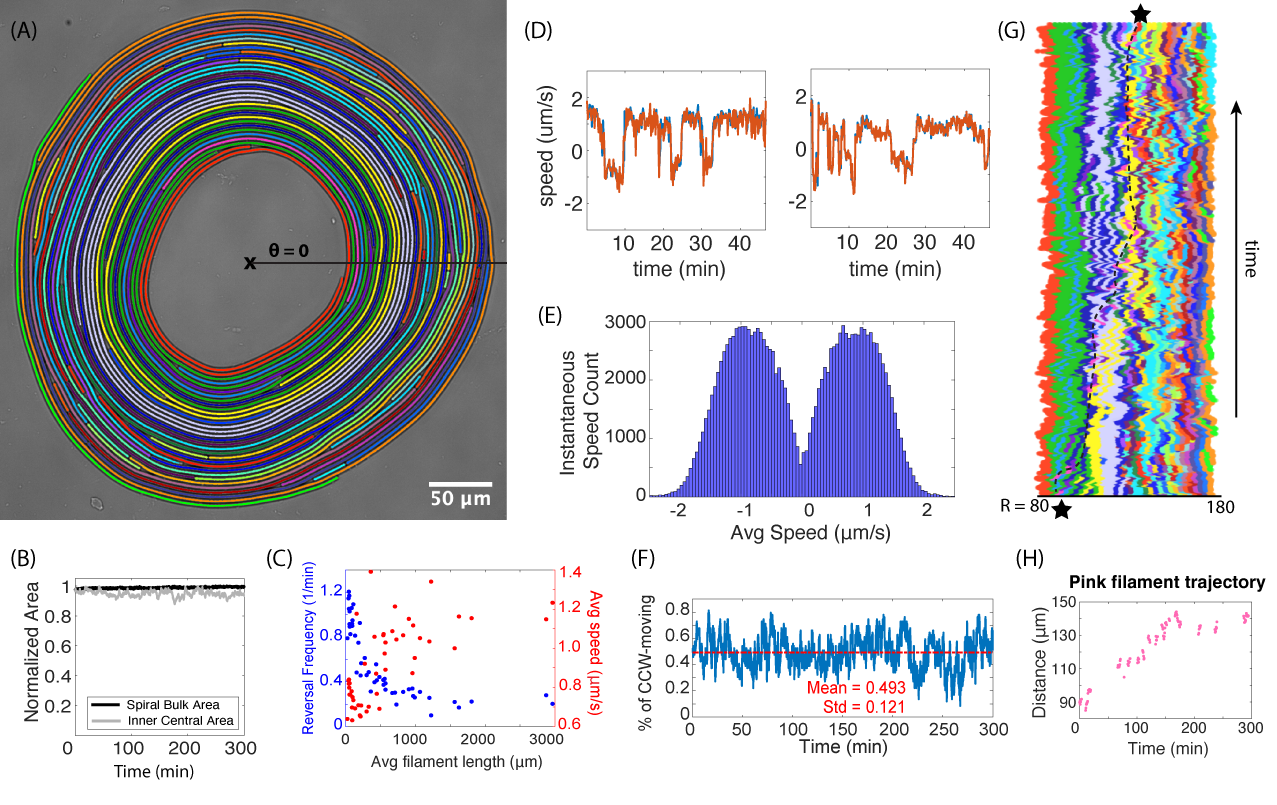}
\caption{\label{Fig2} 
Dynamics and statistics of an active spiral. 
\textbf{(A)} A still image of a spiral with each filament labelled a unique color. A total of 59 filaments are segmented and tracked for 5 hours. ``X'' denotes the approximate spiral center. 
\textbf{(B)} Spiral bulk and inner center size as a function of time shows that the spiral achieves a steady state size.
\textbf{(C)} Average filament speeds and reversal frequencies as a function of filament length. Shorter filaments reverse more frequently than longer filaments, but longer filaments move with higher average speed. 
\textbf{(D)} Example velocity traces of 2 tracked filaments for the first 50 minutes. The orange and blue curves correspond to the linear velocities of each individual tip on a filament. Note the presence of multiple directional reversals.
\textbf{(E)} A histogram of the instantaneous speeds of all tracked filaments across all times shows a symmetric, bimodal distribution with peaks close to $\pm 1 \mu$m/s. 
\textbf{(F)} The number of filaments moving counter-clockwise at any given time fluctuates about a mean of approximately $49 \%$, suggesting that the spiral object as a whole does not favor one direction over another. 
\textbf{(G)} In order to identify material flux along the radial direction, we plot a kymograph of filament identity taken at the slice $\theta = 0$. The horizontal axis represents radial distance (in $\mu$m) from the defect center, while the vertical axis denotes forward evolution in time. Note that multiply-wound filaments will intersect at multiple locations along $R$. By tracking the columns of colors in the kymograph, one can track the internal rearrangement of filaments with respect to the spiral center. As an example, the two black stars denote the initial and final positions of a pink filament that begins close to the spiral center. 
\textbf{(H)} Radial position of the pink filament whenever it crosses $\theta = 0$ as a function of time. Over the course of 5 hours, the filament traverses a radial distance of 52 $\mu$m.
}

\end{figure*}

\begin{figure*}
\includegraphics[width=1\textwidth]{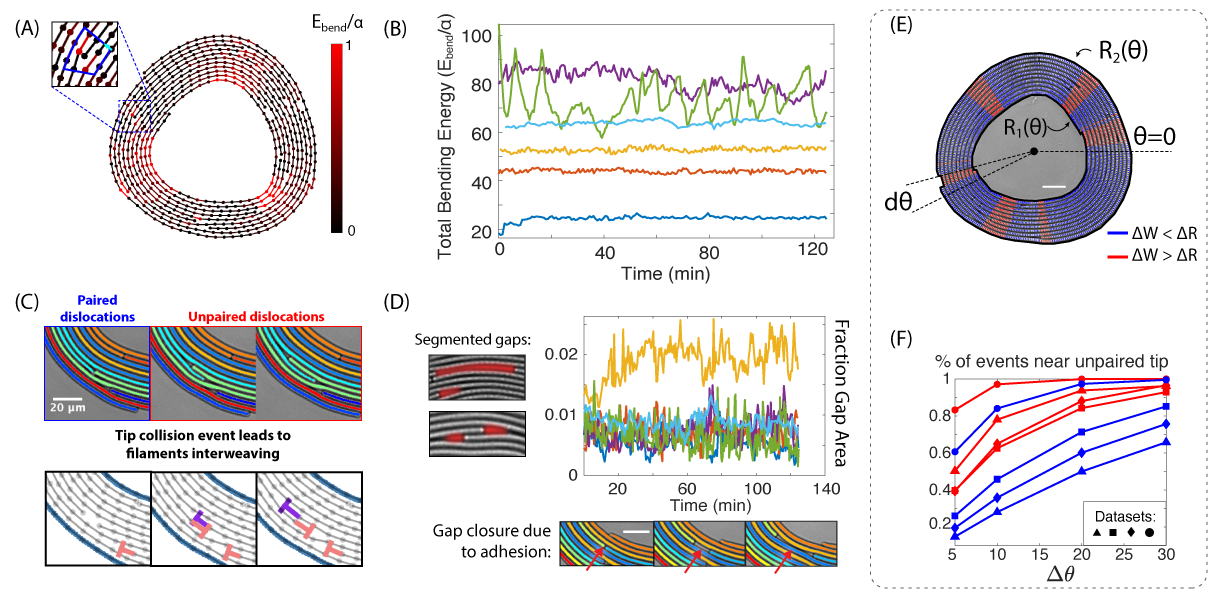}
\caption{\label{energy_fig} 
Tip dynamics and adhesion forces drive shape fluctuations in an active spiral. All unlabelled scale bars show 20$\mu$m.
(A) Individual filaments are skeletonized and the value of $E_{\text{bend}}/\alpha$ is plotted. For ease of visualization, the max value of $E_{\text{bend}}/\alpha$ is set to $\langle E_{\text{bend}}/\alpha \rangle + 3*\sigma_E$, of which 98\% of values fall under. The burgers vector always points radially (inset).
(B) The total bending energy of spiral structures plotted as a function of time reveals that the spirals achieve a constant size about which they fluctuate.
(C) Examples of paired and unpaired dislocations. Note that tips can switch from paired to unpaired states, which result in tips colliding and filaments interweaving.
(D) Gaps in the spiral bulk can be segmented. Their total area as a function of time shows that gaps make up $<3\%$ of the spiral mass. This is a result of adhesion forces in the system, which tend to suppress gap formation (bottom panel).
(E, F) Spiral shape fluctuations are correlated with the presence of dislocations. (E) Sections labelled in red correspond to regions where the change in width $\Delta W(\theta) > \Delta R$ between one frame and the next, while sections labelled in blue correspond to regions where $\Delta W < \Delta R$. 
(F) The fraction of events with a dislocation in a neighborhood $\theta \pm \Delta \theta$. The blue curves denote events where $\Delta W < \Delta R$, while the red curves denote events where $\Delta W > \Delta R$. Each of the different markers correspond to a different tracked spiral. We see from these curves that the presence of dislocations are more strongly correlated with boundary displacement events. 
}
\end{figure*}

\begin{figure*}
\includegraphics[width=\textwidth]{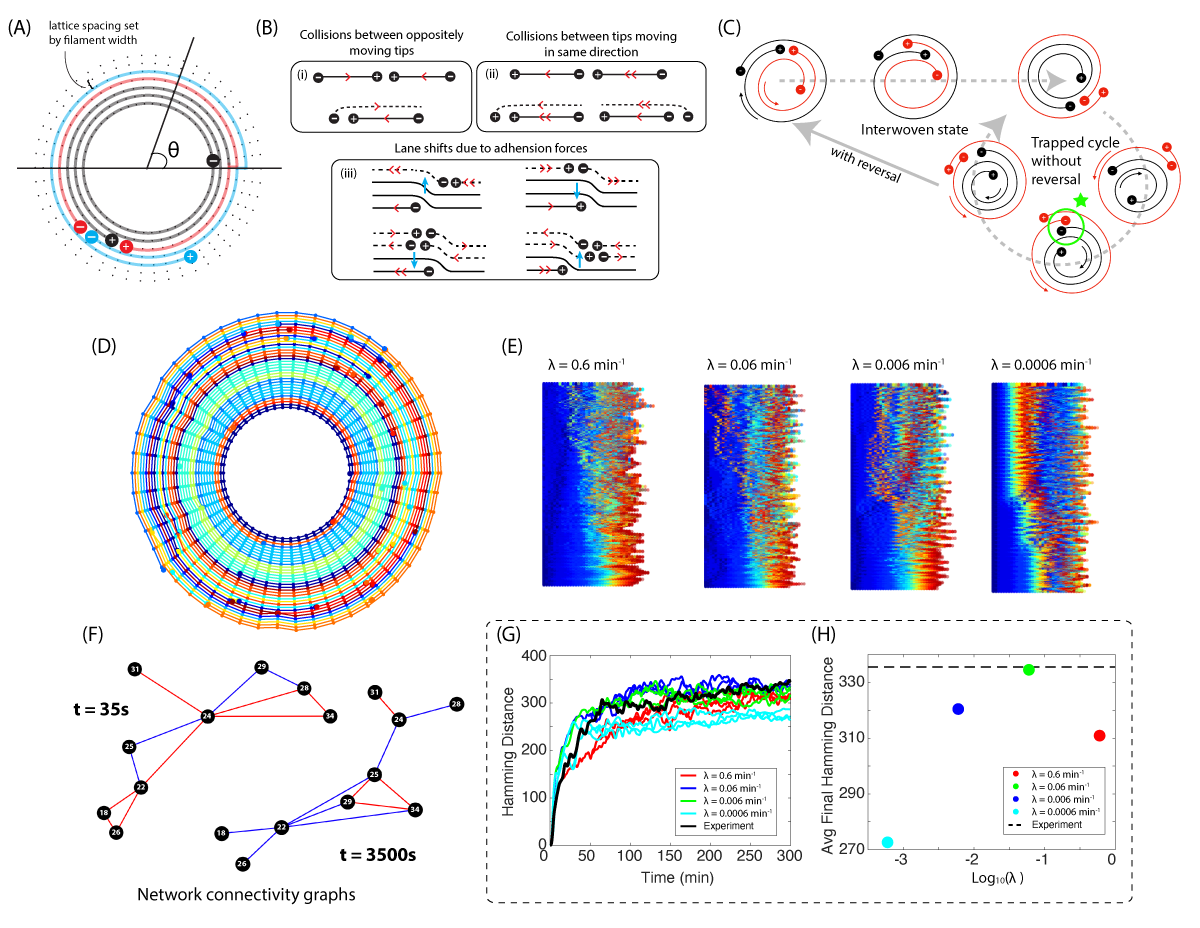}
\caption{\label{sim_material_flux} 
Tip interaction rules on a polar lattice predict material flux.
\textbf{(A)} Filament tips live on a discretized polar coordinate lattice, and interact via tip interaction rules listed in \textbf{(B)}. 
\textbf{(B,i)} Collision events between pairs of tips moving in the opposite direction. Only collision events between two plus tips are stochastic, with either tip having equal probability of being bumped up a channel. In the second case, the minus tip is constrained by geometry to be bumped up.
\textbf{(B,ii)} Collision events between pairs of tips moving in the same direction. In this case, the collision event between the plus and minus tip is stochastic, and the remaining two are deterministic due to geometric constraints.
\textbf{(B,iii)} Lane shifts of tips due to adhesion forces. On the left hand side, a minus tip in a lower channel causes tips above it to either shift up or down a lane depending on whether those tips move over or behind the minus tip. A complementary set of rules for plus tips moving in a lower channel is shown on the right hand side. 
\textbf{(C)} Cartoon schematic showing that in the absence of reversal, two oppositely moving filaments (with the same geometric chirality) will reach a steady state configuration.
\textbf{(D)} Simulation of tip interaction rules at time 0, designed to mimic the experimental configuration in Fig. 2A. 46 of the original 59 filaments are included in the simulation (the shortest 13 filaments with lengths $< 50 \mu$m are excluded). 
\textbf{(E)} Filament identity kymographs taken at $\theta = 0$ for four different reversal frequencies.
\textbf{(F)} Each spiral snapshot can be converted into a connectivity graph, where nodes represent filaments and edges exist between filaments that touch due to being in adjacent channels. An edge can only exist if the two filaments touch for an angle of at least $14^{\circ}$.
\textbf{(G)} Hamming distance as a function of time for various reversal dynamics.
\textbf{(H)} Average steady state values of hamming distance show an optimum value with respect to reversal frequency.
}
\end{figure*}

\begin{figure*}
\includegraphics[width=\textwidth]{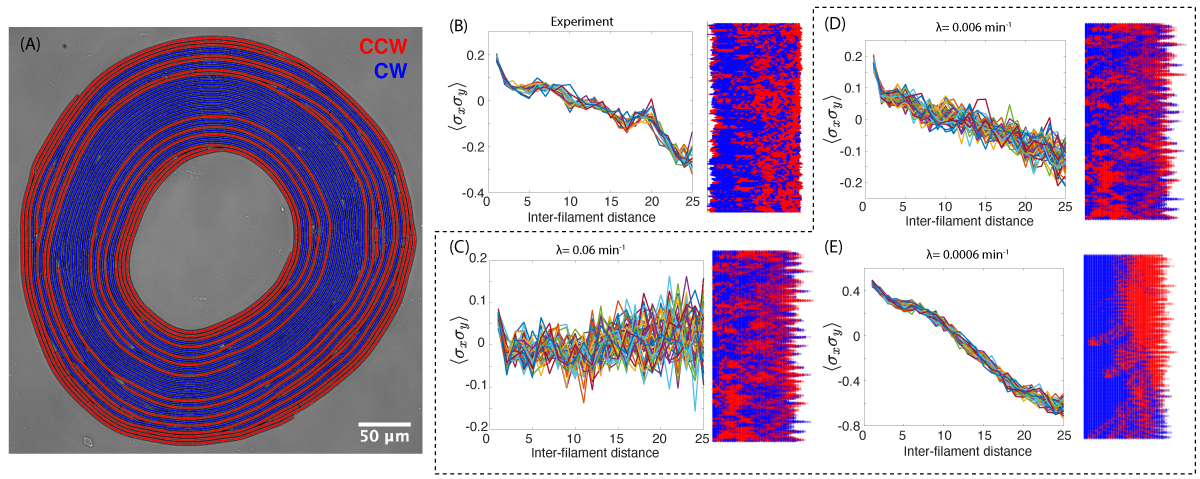}
\caption{\label{sim_dir_corr} 
Filament direction correlation plots.
\textbf{(A)} Snapshot of the spiral where each filament is labelled by direction of motion (Red = CCW; Blue = CW). 
\textbf{(B-E)} Left: Direction correlation as a function of inter-filament distance (x-axis plotted in multiples of filament body-widths). Right: direction kymograph taken at $\theta = 0$.
Comparing experimental data in \textbf{(B)} to simulation results \textbf{(C-E)} shows that a reversal frequency of $\lambda = $ 0.006 min$^{-1}$ most closely reproduces experimental results. However, this value is two orders of magnitude smaller than the average measured reversal rate (Fig. 2C, blue), suggesting that biological feedback mechanisms are at play which regulate filament motility. These higher order effects are not captured by our simplistic model. 
}
\end{figure*}

\begin{figure*}
\includegraphics[width=\textwidth]{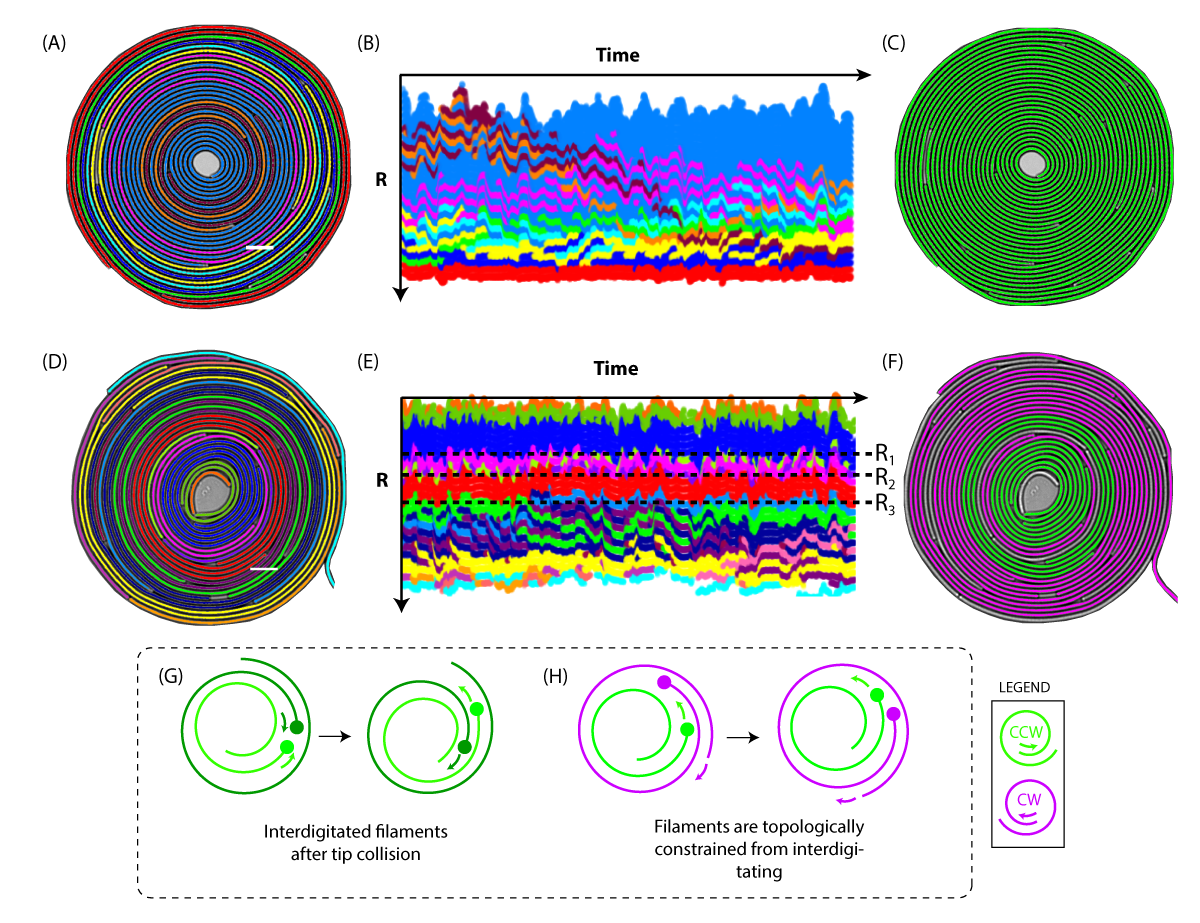}
\caption{\label{top_trap} 
Topological traps formed by filaments of opposite geometric chirality. All unlabelled scale bars show 20$\mu$m.
Two labelled spirals \textbf{(A,D)} and their respective kymographs in \textbf{(B,E)}. The fact that material flux is limited to the regions between the dotted lines in \textbf{(E)} is explained by the geometric chiralities of the filaments within each spiral. 
\textbf{(C,F)} The two spirals now labelled by geometric chirality, defined as the physical winding of the filament as you trace its length from the inner tip (see LEGEND inset). The spiral in \textbf{(A)} is uniformly composed of CCW wound filaments, whereas the spiral in \textbf{(D)} has alternating regions of CCW and CW filaments.  \textbf{(G,H)} Schematics showing how geometric chirality limits radial flux. 
}
\end{figure*}

\newpage 
\bibliography{references2.bib}


\renewcommand\thefigure{S\arabic{figure}} 
\setcounter{equation}{0}
\setcounter{figure}{0}
\newpage 

{\centering \section{Supplementary Information}}

\subsection{I. Organism and culture}

The organism \textit{Oscillatoria Sp.} was purchased from carolina.com and cultured in glass petri dishes in an incubator set to a temperature of 22$^\circ$C. A single fluorescent light tube was used to illuminate the cultures, and set to a 12 hr day-night cycle. Alga-gro Freshwater Medium, also purchased from carolina, was used to culture the cyanobacteria.

\subsection{II. Timelapse set up}

The experimental set up used for imaging consists of \textit{Oscillatoria} filaments sandwiched between a 2\% agarose gel and coverslip (see Fig. \ref{Fig1}, main text). First, one side of a 65$\mu$L gene frame (1.5 x 1.6 cm in length and width) is stuck to the surface of a glass side. Liquid agarose is then poured into the gene frame and compressed with a coverslip to ensure that it dries as a flat surface. Note that this first coverslip is removable, as the plastic cover of the free surface of the gene frame is not yet removed.

When preparing the filaments for plating on the gel, a dense clump is first placed on a glass slide. A pipette tip is then used to break up dense filament clumps, creating a more dilute sample with which to plate the gel. It is imperative that before the coverslip is set permanently on the gene frame, that enough water is allowed to evaporate. If too much water is trapped between the gel and coverslip, the filaments in between will not gain enough traction to glide. After waiting a few minutes for some of the water to evaporate, the plastic cover of the gene frame is removed and a fresh coverslip is laid carefully down. The entire system is thus sealed, preventing subsequent evaporation. 

Imaging is done under uniform, low-light conditions using a 20x objective lens.

\subsection{III. Segmentation and Tracking}
Segmentation and tracking are two separate algorithmic steps needed to track the motion of individual filaments. Segmentation is essentially an image classification problem, and entails identifying filament edges in every frame. Tracking is then built on top of the segmented images, where individual segments are mapped to their corresponding identities in consecutive frames.

\noindent \underline{Segmentation}\\
Segmentation was performed using a combination of Matlab's image processing toolbox and a convolution-neural-network architecture called the UNET [1]. Specifically, the repository used was downloaded from https://github.com/zhixuhao/unet [2]. 
Classical image processing operations were first performed in matlab, which then provided training images for the UNET. Though Matlab's segmentation was decent, it was not robust against non-uniform imaging conditions or fluctuations in the focal plane, which frequently occur during long timelapses. 
The remaining segmentation errors from the UNET outputs were manually corrected for in Adobe Illustrator.

\noindent \underline{Tracking}\\
Given the segmentation, filaments in each frame were tracked using a greedy algorithm which maximized the score: $s^t_{ij} = \frac{\text{overlap}(f_i^t, f_j^{t+1})}{\Delta L_{ij}}$. Here, $f_i^t$ denotes the segmented filament $i$ in frame $t$, while $f_j^{t+1}$ denotes a segmented filament $j$ in the next frame, $t+1$. The greater the area of overlap between the two filaments, the higher the score. On the other hand, since filaments do not change in length from frame to frame, $\Delta L_{ij}$ penalizes assigning filaments with large discrepancies in length.

\subsection{IV. Individual Tip Dynamics}
In this section we discuss a subtlety of individual tip dynamics which plays a role both in understanding the spiral system and in implementing our simulations. We have shown that filament lengths grow negligibly over the course of our experiments, and so the constancy of filament lengths is an important constraint.

Intuitively, if one imagines an unwound filament gliding on a gel, one would expect that the two filament tips move at equal linear velocities. This implies then, in the case of a wound filament, that the angular velocities of the tips vary as a function of tip radial position (specifically, $\omega \sim 1/R$), where $v = 1 \mu$m/s is the average measured tip linear velocity (Fig. \ref{Fig2}E, main text). Indeed, plotting the instantaneous angular velocities of the individual tips as a function of R shows that the data on average follows this trend (Fig. \ref{03052020-ang_vs_R}B,\ref{06302020xy1-ang_vs_R}B). 

\begin{figure*}
\includegraphics[width=\textwidth]{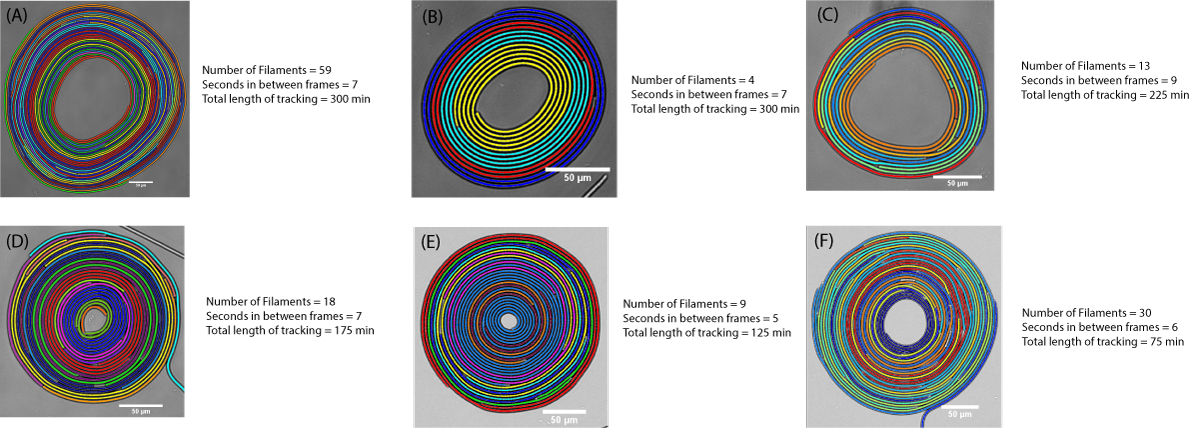}
\caption{\label{spiral_stats} 
(A-F) 6 tracked spirals and their descriptions.
}
\end{figure*}

\begin{figure*}
\includegraphics[width=0.75\textwidth]{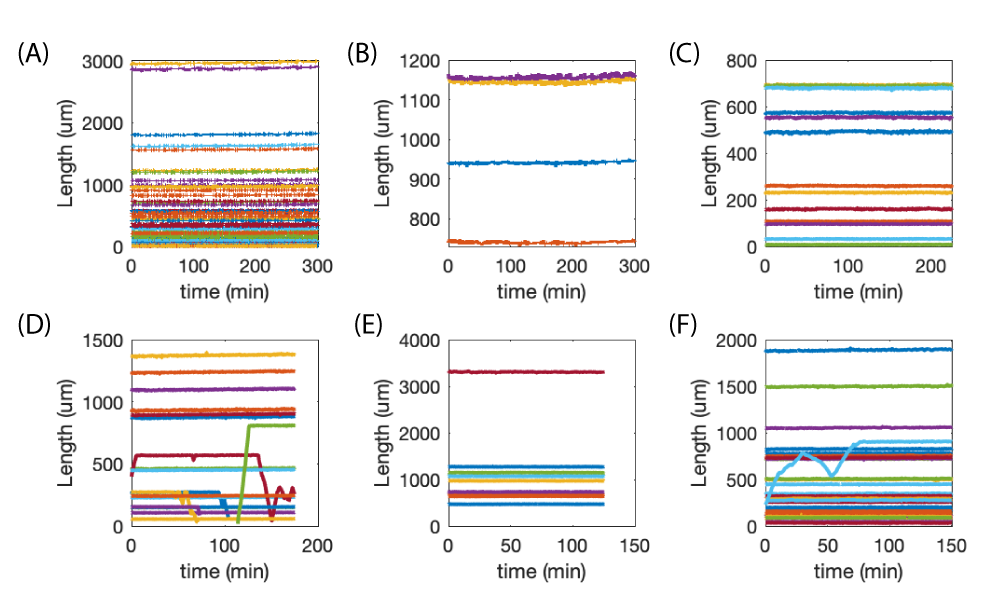}
\caption{\label{lengths_vs_time} Filaments experience minimal growth under low light conditions.
(A-F) Lengths vs. time of each tracked filament for the 6 spiral structures, with the lettering in this figure corresponding to the lettering in Fig. S1. The perceived length changes in (D) and (F) are due to filaments gliding in and out of the field of view.
}
\end{figure*}

\begin{figure*}
\includegraphics[width=\textwidth]{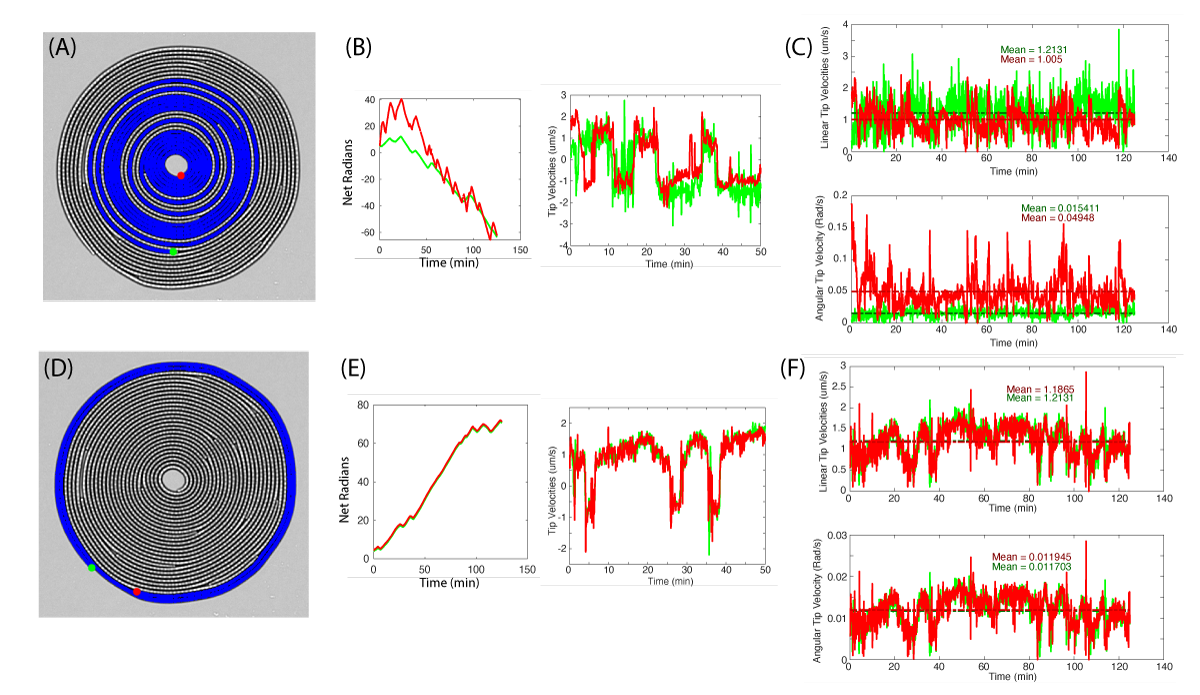}
\caption{\label{tips_moving_opp_dir} Filament tips can move in opposite directions for brief periods of time.
(A) The inner-most, tightly wound filament depicted in blue, with inner and outer tips labelled in red and green, respectively. The labelled filament is 3305 $\mu m$ long. 
(B) The net angle traced by each tip (left) and the tip velocities as a function of time (right). Note that the two tips move in different directions at times $t < 5$ min. 
(C) The average angular and linear speeds of each tip shows that there is a large discrepancy in the angular speeds while the linear speeds are approximately the same. The green (outside) tip moves at an angular speed that is only 30\% that of the red tip. 
(D) The outer most tip is depicted in blue with the inner and outer tips labelled accordingly.
(E) Same as in (B). Note that the two tips move much more synchronously than in (B). 
(F) The average linear and angular velocities show very little discrepancy, owing to the fact that the two tips sit at approximately the same radial distance from the spiral center.
}
\end{figure*}

\begin{figure*}
\includegraphics[width=0.8\textwidth]{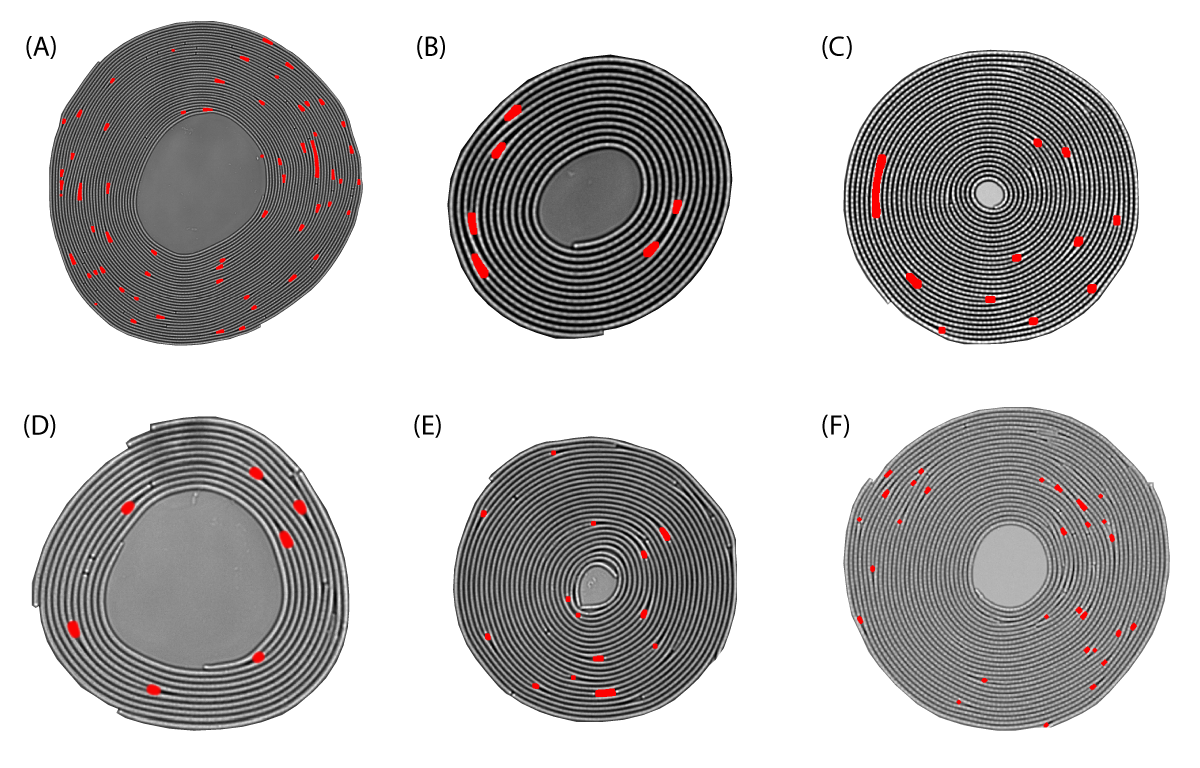}
\caption{\label{gaps} 
(A-F) Gaps segmented from each of the 6 tracked spirals.
}
\end{figure*}

\begin{figure*}
\includegraphics[width=0.8\textwidth]{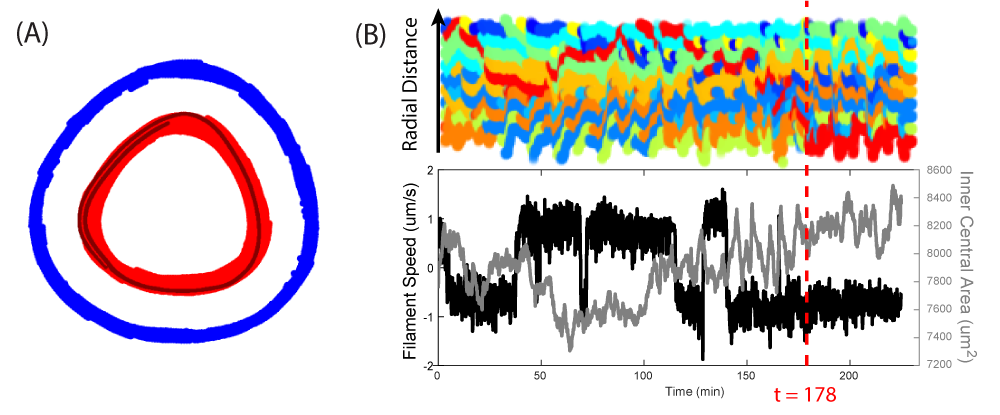}
\caption{\label{radial_slippage} 
Radial slippage of filaments in a spiral helps drive spiral shape relaxation (the spiral maintains steady state size).
(A) The red and blue regions denote the locations of the inner and outer boundaries $R_1$ and $R_2$, respectively, over the course of approximately 50 minutes. The dark red line overlaid in the red region shows a snapshot of the innermost filament. Hence, the boundaries are capable of radial slippage of distances thicker than a single filament width. 
(B) (Top panel) Kymograph of filament identities at $\theta = 0$ over time. Note that from $t > 178$ min, the inner most filament remains constant (labelled bright red). (Bottom panel) From $t > 178$ min, the inner most (bright red) filament continues to wind clockwise, yet the inner central area of the spiral does not net decrease. 
}
\end{figure*}

\begin{figure*}
\includegraphics[width=0.5\textwidth]{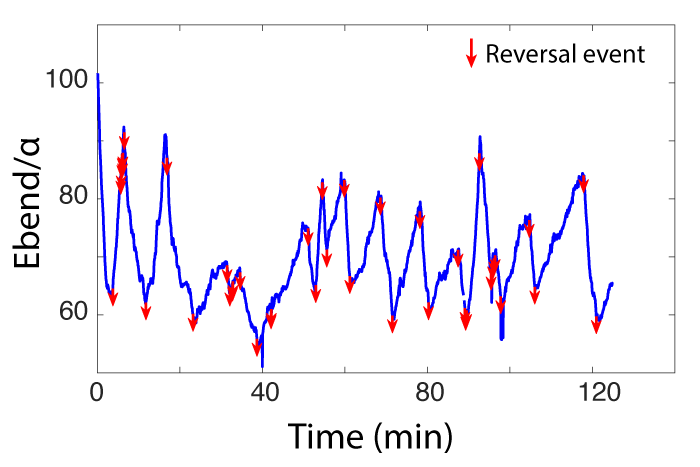}
\caption{\label{reversal_arrows} 
Oscillations in spiral (Fig. S1(E) data set) bending energy corresponds to reversal events of the innermost filament.
}
\end{figure*}

\begin{figure*}
\includegraphics[width=\textwidth]{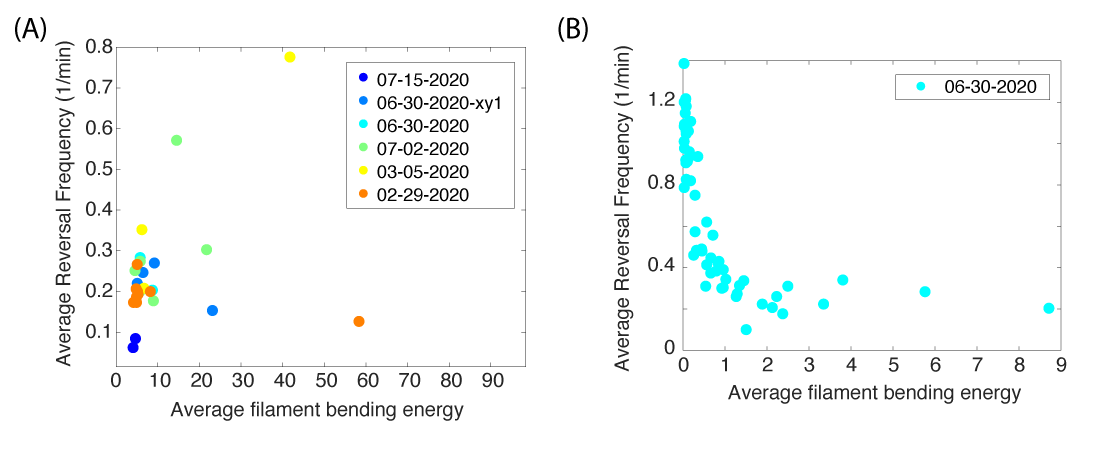}
\caption{\label{reversal_vs_ebend} 
At high bending energies, filament reversal frequency appears positively correlated with bending energy. 
(A) Average filament bending energy vs. average reversal frequency are plotted for all filaments with average bending energy $E_\text{bend}/\alpha > 4$. There appears to be a trend where the filaments reverse more frequently if more tightly wound. 
(B) The same curve is plotted for all filaments in the spiral of Fig. S1(A), which is also the least tightly wound spiral. In this case there is an inverse relationship between bending energy and reversal frequency, and the shape of this curve closely mimics that of Fig. 2C, blue. In other words, shorter filaments tend to have less bending energy and reverse the most frequently.
}
\end{figure*}

\begin{figure*}
\includegraphics[width=0.5\textwidth]{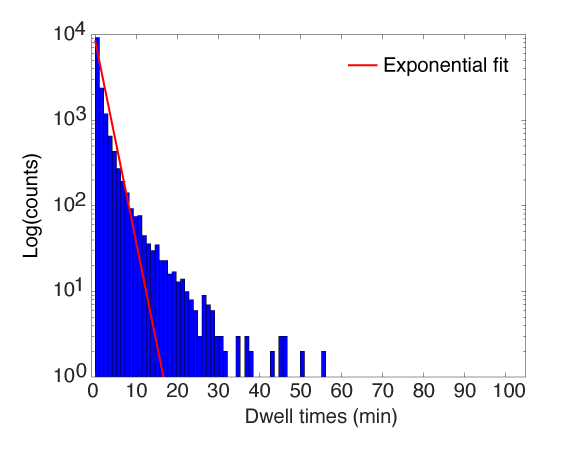}
\caption{\label{heavytail_dwelltime} 
The dwell times (time in between reversal events) for all filaments across all tracked spirals. Y-axis is in log-scale. The distribution shows a heavy tail, meaning there is an enhanced probability of long runs.
}
\end{figure*}

\begin{figure*}
\includegraphics[width=\textwidth]{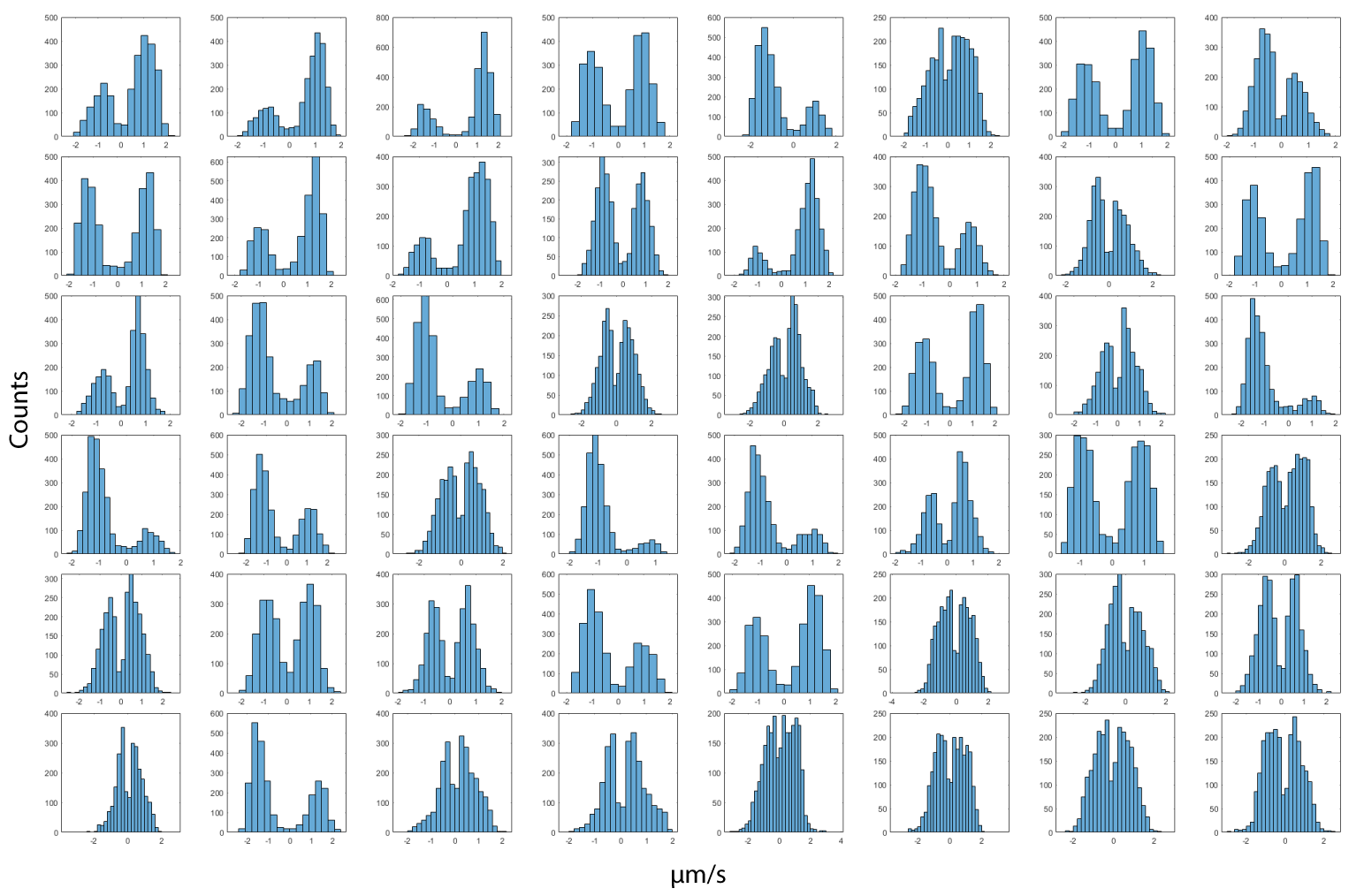}
\caption{\label{063030_spd_histograms} 
Speed histograms for individual filaments in the Fig. S1(A) dataset are plotted. Note that some velocity distributions show significant skew in one direction versus another. 
}
\end{figure*}

\begin{figure*}
\includegraphics[width=\textwidth]{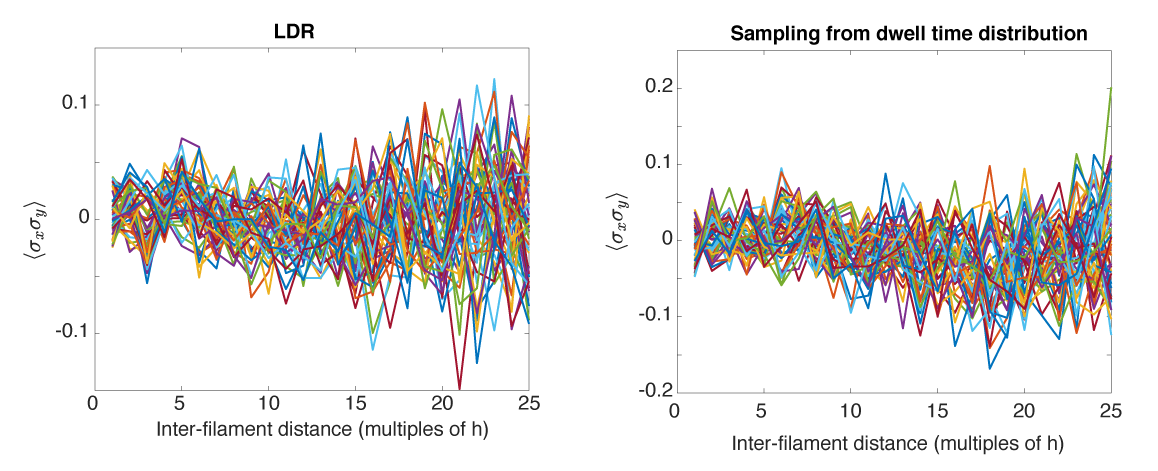}
\caption{\label{LDR_dwell_dir_corr} 
The directionality correlation plotted for simulations run by (A) introducing empirically measured length-dependent rates of reversal, and (B) by sampling dwell times from the empirically measured dwell-time distribution. Neither situation is able to reproduce the directionality-correlation observed experimentally.
}
\end{figure*}

\begin{figure*}
\includegraphics[width=0.8\textwidth]{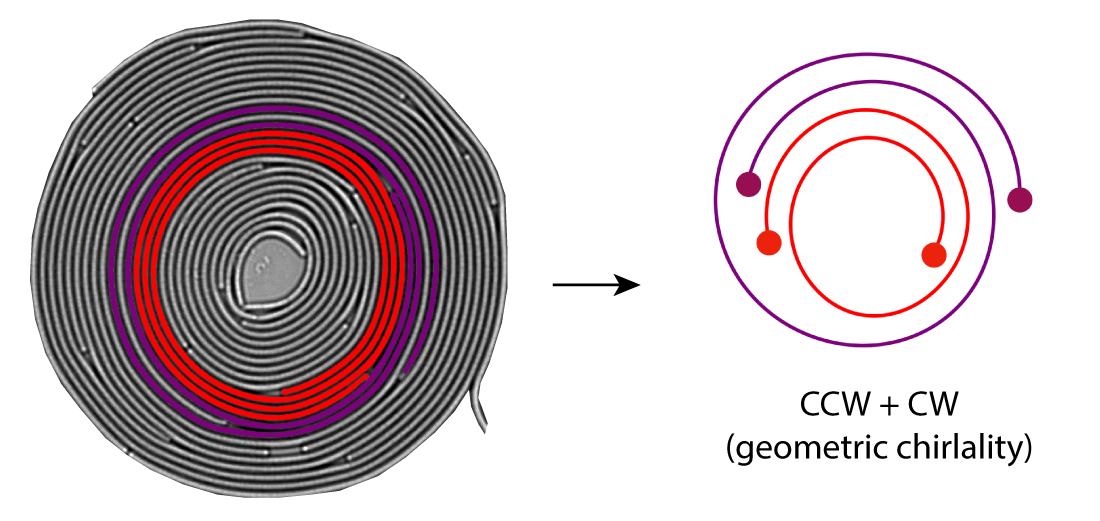}
\caption{\label{07022020-fil9_fil11} 
The two filaments that form the boundary of a topological trap are highlighted. The inner (red) filament winds geometrically CCW, while the outer (purple) filament winds geometrically CW, together forming a barrier across which material cannot flow.
}
\end{figure*}

\begin{figure*}
\includegraphics[width=\textwidth]{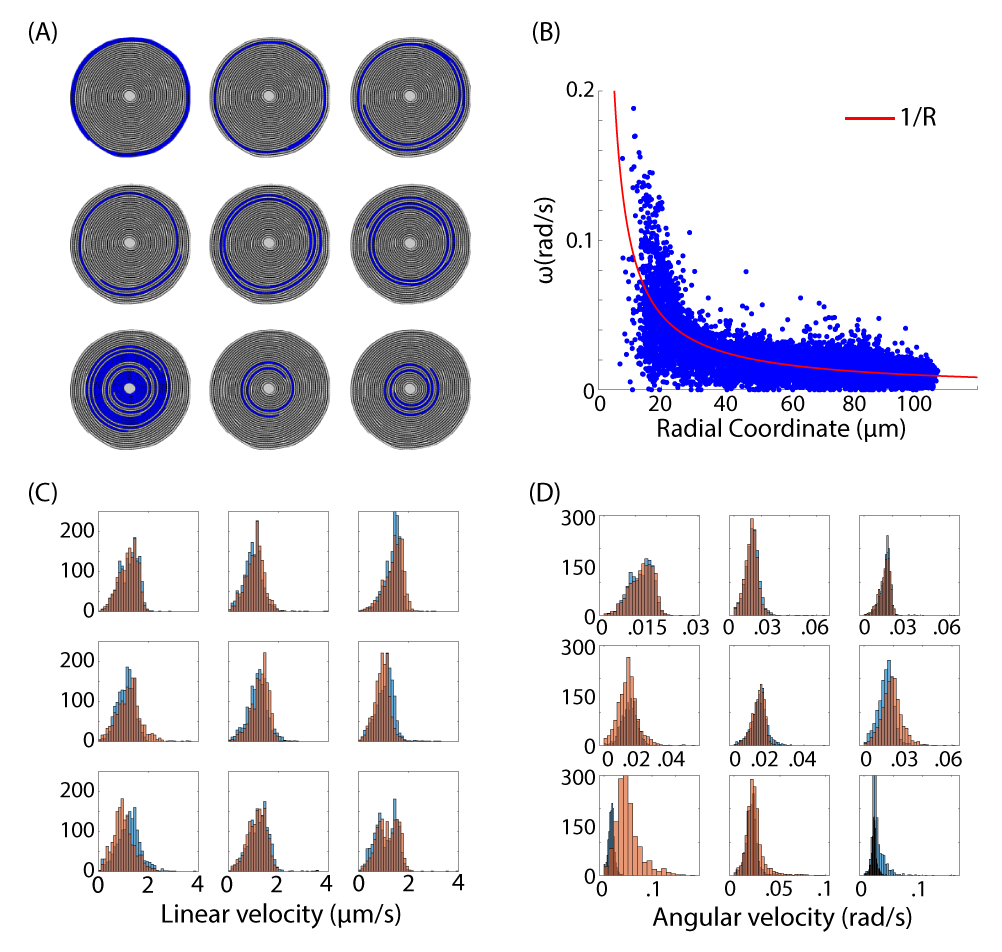}
\caption{\label{03052020-ang_vs_R} 
Individual tip dynamics for the Fig. S1(E) data set show that tips on the same filament move at the same \textit{linear} velocity. (A) Each of the 9 filaments within the spiral are individually highlighted. (B) The instantaneous angular tip velocities as a function of radial coordinate shows that it roughly follows $\omega \sim 1/R$, where $v = 1\mu$m/s is the average linear velocity of the tips. (C) Histogram of tip linear velocities for each of the 9 filaments (each panel corresponds to one of the filaments in (A)). The histograms for tips on the same filament are approximately overlapping. (D) Histogram of tip angular velocities show that angular velocities can have very different distributions for tips on the same filament.
}
\end{figure*}

\begin{figure*}
\includegraphics[width=\textwidth]{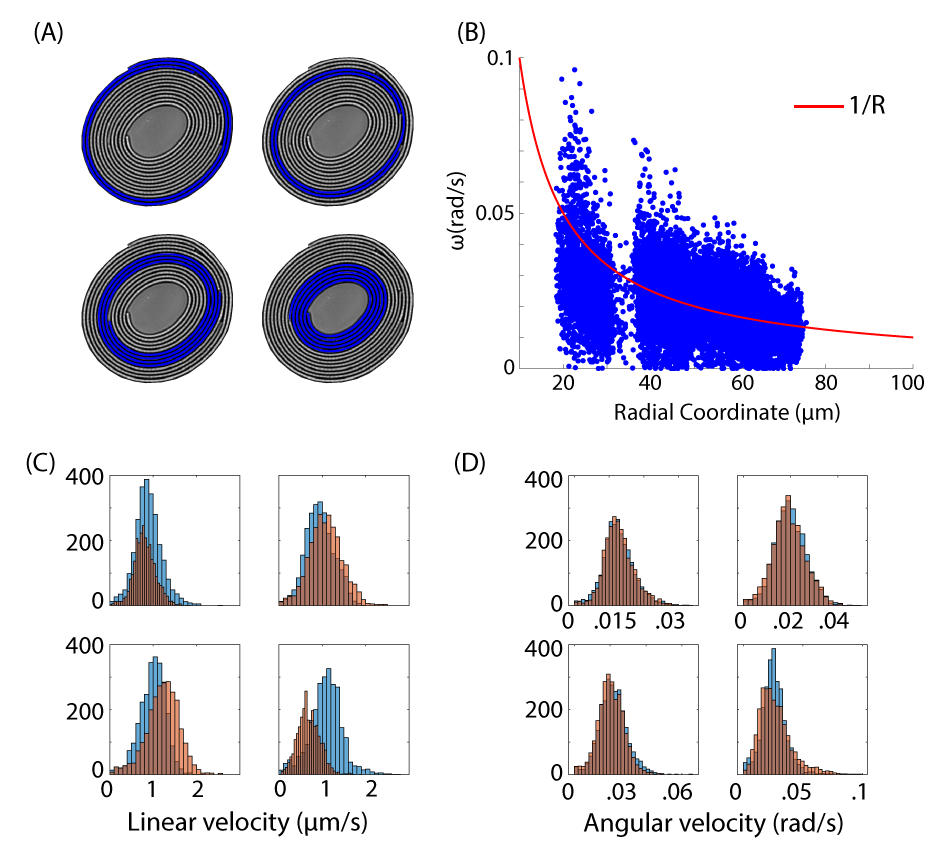}
\caption{\label{06302020xy1-ang_vs_R} 
Individual tip dynamics for the Fig. S1(B) data set show that, in contrast to Fig. S12, tips on the same filament tend to move at the same \textit{angular} velocity. (A) Each of the 4 filaments within the spiral are individually highlighted. (B) The instantaneous angular tip velocities as a function of radial coordinate shows that it roughly follows $\omega \sim 1/R$, where $v = 1\mu$m/s is the average linear velocity of the tips. (C) Histogram of tip linear velocities shows that tips on the same filament can have different distributions, so to maintain a constant angular velocity. (D) Histogram of tip angular velocities show that tips on the same filament have overlapping distributions.
}
\end{figure*}

If we return to the filament length constraint, however, this intuition breaks down. 
Take, for example, the innermost yellow filament of data set 06-30-2020-xy1 (Fig. \ref{spiral_stats}B). This filament has winding number = 4, and does not change radial position throughout the entire time lapse (Movie S3). Suppose the two tips move at constant linear velocity $v$, then this would imply that $\omega_i = V/R_i > \omega_o = V/R_o$, where $\omega_i$ and $\omega_o$ denote the inner and outer angular tip velocities, respectively. However, the yellow filament does not change radial position and moreover, tracking reveals that the two tips remain at the same radial coordinate even as the filament traverses multiples of $2\pi$. Then $\omega_i > \omega_o$ would mean that the filament is increasing in length as it winds! This contradiction suggests that the two filament tips must instead be moving at constant angular velocities, which is corroborated by the data in Fig. \ref{06302020xy1-ang_vs_R}C, D (bottom right hand panel). 
Hence, this teaches us that though on average tips move at linear velocities of ~1 um/s, individual tips on a wound filament can have discrepancies in linear speeds to satisfy length constraint. These discrepancies may arise from radial slippage (so that the velocity is not always tangential) and adhesion fores (that allow a wound filament to glide as though it is rotating like a solid body).

\subsection{V. Simulation Details}

In our simulation framework, we model active spirals as a system of self-propelled tips on a polar coordinate lattice. Each tip is thereby constrained to move on circular tracks, and traverse in either the CW or CCW directions. As explained in the main text, tips can shift up or down into neighboring lanes according to the rules depicted in Fig. \ref{sim_material_flux} (main text). There is an additional subtly owing to how the tips move in the \textit{angular} coordinate. 

\begin{figure*}
\includegraphics[width=\textwidth]{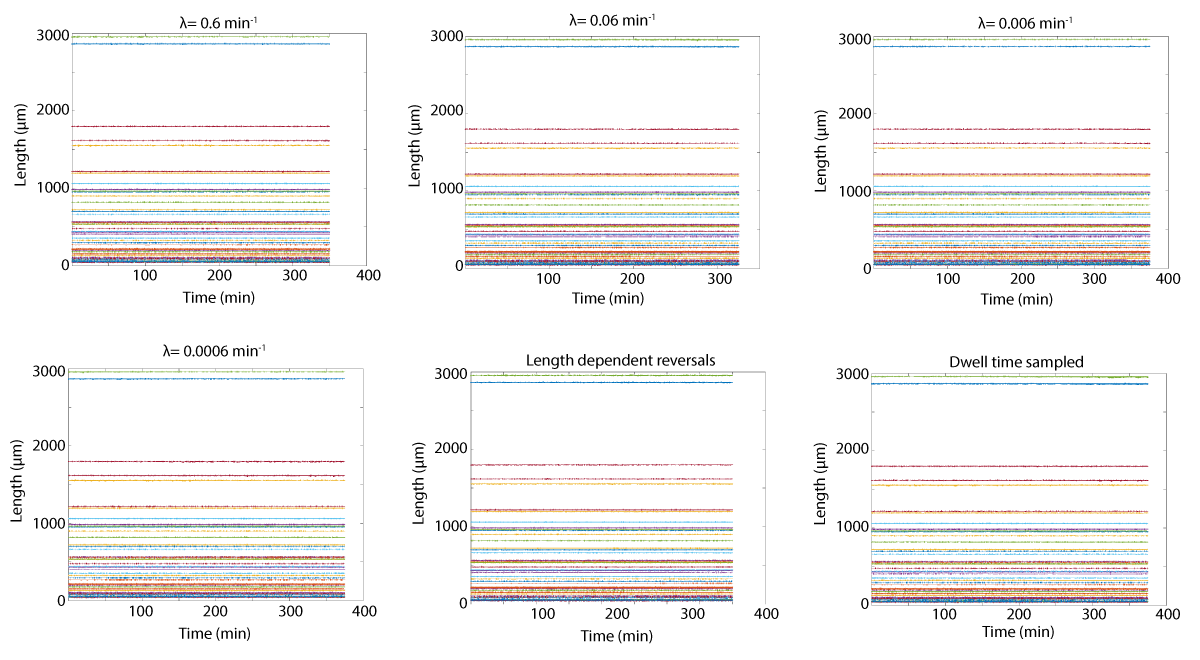}
\caption{\label{sim_lengths_vs_time} 
Filament lengths remain constant for each of the 6 simulation set ups. 
}
\end{figure*}

The plus (leading) and minus (trailing) tips of a filament must move in a way that keeps the filament length constant, as we work in the limit of 0 growth. However, the plus and minus tips of a filament need not lie in the same channel (i.e. they can have different radial coordinates). If the two tips move at the same linear velocity, this would imply different angular velocities, and hence a change in filament length (see section IV above for a more detailed explanation). Hence, there is ambiguity in terms of how to define the instantaneous velocities at each tip. The simplest method to satisfy the filament length constraint, and the one implemented in our simulations, is to update the angular coordinate of the plus tips according to $\Delta \theta = (v/R) \Delta t$, where $v = 1 \ \mu$m/s is the linear velocity of the filament and $R$ the radial coordinate, and update the minus tip such that $\Delta L$ = 0, with $L$ the filament length. In other words, the minus tip simply ``catches up'' to the plus tip, keeping the filament length unchanged. Fig. \ref{sim_lengths_vs_time} shows that filament lengths indeed remain constant over the course of each of the simulations.

Finally, the radial coordinate of the tips are updated according to the rules prescribed in Fig. \ref{sim_material_flux}B of the main text.

\newpage

\subsection{VI. Supplementary figures}

\clearpage
\noindent \textbf{Movie 1:}
This movie shows the formation of spirals after initial plating on an agarose gel. 

\noindent\textbf{Movies 2-7:}
Movies 2-7 show 6 segmented and tracked spirals. Every filament is individually labelled within the spirals.

\noindent\textbf{Movie 8:}
This movie shows the two tips of the long, innermost filament in Fig. S3A moving in opposite directions. The long filament is labelled in blue on the left hand panel, while the dynamics of the two tips (denoted by red and green dots) are individually shown in the right hand panel.

\noindent\textbf{Movie 9:}
Radial flux of filaments in a spiral (data set Fig. S1(C)). The left hand side of the movie shows the tracked spiral, while the right hand side shows how the filament kymograph (taken at $\theta = 0$) evolves over time.

\noindent\textbf{Movie 10:}
Innermost filament of spiral (data set Fig. S1(C)) continues winding inward without leading to net decrease in size of spiral center.

\noindent\textbf{Movie 11:}
Spiral (data set Fig. S1(C)) with all dislocations (unpaired tips) labelled. This is to illustrate that a spiral can be modelled as a system of active dislocations with climb dynamics (movement of dislocation perpendicular to its burgers vector).

\noindent\textbf{Movie 12:}
Sample simulation video, with filament configurations modelled after data set Fig. S1(A). Only filaments with length $>50\mu$m are simulated, which incorporates 46 of the 59 filaments. The reversal frequency in this video is set to $\lambda = 0.06$ min$^{-1}$.

\noindent\textbf{Movie 13:}
Spiral in data set Fig. S1(D) is shown with all filaments labelled by identity (left hand side) and all filaments labelled by geometric chirality (right hand side). The boundaries where two neighboring filaments have opposite geometric chirality form topological traps, which material cannot move past. 

\noindent\textbf{Movie 14:}
The innermost filament of data set Fig. S1(B) and the dynamics of its two tips shown side-by-side. The two tips move at constant angular velocity, and hence different linear velocities. 

\subsection{SI References}
\begin{itemize}
    \item[[1]] O. Ronneberger, P. Fischer, and T. Brox, “U-net: Convolutional networks for biomedical image
segmentation,” in International Conference on Medical image computing and computer-assisted
intervention (Springer, 2015) pp. 234–241.
    \item[[2]] zhixuhao, “Implementation of deep learning framework unet, using keras,” \\ https://github.com/zhixuhao/unet (2019).
\end{itemize}

\end{document}